\documentclass{aa} 

\usepackage{graphicx}
\usepackage{indentfirst} 
\usepackage{txfonts}
\usepackage[utf8]{inputenc}
\usepackage[T1]{fontenc}

\emergencystretch=\maxdimen
\usepackage{dcolumn} 
\usepackage{textgreek}
\usepackage{morefloats}
\usepackage{pdflscape}
\usepackage{afterpage}
\usepackage{graphicx}
\usepackage{changepage}
\usepackage{longtable}
\usepackage{threeparttable}
\usepackage{multicol}
\usepackage{multirow}
\usepackage{rotating}
\usepackage{hyperref}
\usepackage{float}
\usepackage{txfonts}
\usepackage{siunitx}
\usepackage[Symbolsmallscale]{upgreek}
\usepackage{fancyvrb,cprotect}

\usepackage{etoolbox}

\usepackage{cleveref}
\crefname{section}{§}{§§}
\Crefname{section}{§}{§§}

\usepackage{color, colortbl}
\definecolor{Gray}{gray}{0.9}
\definecolor{Green}{rgb}{0,1,0}
\definecolor{Red}{rgb}{1,0,0}
\usepackage[table]{xcolor}

\usepackage{hyperref}
\usepackage{xcolor}
\hypersetup{
  colorlinks   = true, 
  urlcolor     = black, 
  linkcolor    = blue, 
  citecolor    = blue 
}
\usepackage[all]{hypcap} 

\usepackage{comment}

\begin{document}

  \title{Abell\,1430: A merging cluster \\ with exceptional diffuse radio emission}
  \titlerunning{Radio emission in Abell\,1430}

  \author{
    M. Hoeft \inst{1} \and
    C. Dumba \inst{2,1} \and
    A. Drabent \inst{1} \and
    K. Rajpurohit \inst{3,4,1} \and
    M. Rossetti  \inst{5} \and
    S.~E. Nuza \inst{6,7} \and
    R.~J. van Weeren \inst{8} \and 
    H. Meusinger \inst{1} \and
    A. Botteon \inst{8} \and
    G. Brunetti \inst{4} \and
    T.~W. Shimwell \inst{9,8} \and
    R. Cassano \inst{4} \and
    M. Brüggen  \inst{10}   \and
    H. J. A. Röttgering \inst{8} \and
    F. Gastaldello  \inst{5} \and
    L. Lovisari \inst{11,12} \and
    G. Yepes \inst{13} \and
    F. Andrade-Santos  \inst{11} \and
    D. Eckert \inst{14} 
    }

  \institute{ 
    Thüringer Landessternwarte, Sternwarte 5, 07778 Tautenburg, Germany\\
    \email{hoeft@tls-tautenburg.de} 
    \and
    Mbarara University of Science \& Technology, PO Box 1410 Mbarara, Uganda
    \and
    Dipartimento di Fisica e Astronomia, Universit\'a di Bologna, via P. Gobetti 93/2, 40129, Bologna, Italy 
    \and
    INAF-Istituto di Radio Astronomia, Via Gobetti 101, 40129, Bologna, Italy
    \and
    IASF-Milano, INAF, via A. Corti 12, 20133 Milano, Italy
    \and
    Instituto de Astronomía y Física del Espacio (IAFE, CONICET-UBA), CC 67, Suc. 28, 1428 Buenos Aires, Argentina
    \and
    Facultad de Ciencias Exactas y Naturales (FCEyN), Universidad de Buenos Aires (UBA), Buenos Aires, Argentina
    \and 
    Leiden Observatory, Leiden University, PO Box 9513, NL-2300 RA Leiden, The Netherlands
    \and
    ASTRON, the Netherlands Institute for Radio Astronomy, Postbus 2, 7990 AA, Dwingeloo, The Netherlands
    \and
    University of Hamburg, Hamburger Sternwarte, Gojenbergsweg 112, 21029 Hamburg, Germany
    \and
    INAF - Osservatorio di Astrofisica e Scienza dello Spazio di Bologna, via Piero Gobetti 93/3, I-40129 Bologna, Italia
    \and
    Center for Astrophysics $|$ Harvard $\&$ Smithsonian, 60 Garden Street, Cambridge, MA 02138, USA
    \and
    Departamento de Física Teórica and CIAFF, Módulo 8, Facultad de Ciencias, Universidad Autónoma de Madrid, E-28049 Cantoblanco, Madrid, Spain
    \and
    Department of Astronomy, University of Geneva, ch. d'Ecogia 16, 1290 Versoix, Switzerland
    }

   \date{Received XXX; accepted YYY}

   \abstract{
     % Context
     Diffuse radio emission has been found in many galaxy clusters, predominantly in massive systems which are in the state of merging. The radio emission can usually be classified as relic or halo emission, which are believed to be related to merger shocks or volume-filling turbulence, respectively. Recent observations have revealed radio bridges for some pairs of very close galaxy clusters. The mechanisms that may allow to explain the high specific density of relativistic electrons, necessary to explain the radio luminosity of these bridge regions, are poorly explored until now. 
     }{
     % Aims
     When inspecting the first data release of the LOFAR Two-Metre Sky Survey (LoTSS), we discovered diffuse radio emission in the galaxy cluster Abell\,1430. Here, we aim at determining the dynamical state of the cluster and characterising the diffuse radio emission.
     }{
     % Methods
     We analyse the LoTSS data in detail and complement it with recent Karl G. Jansky Very Large Array observations in the L-band. To study the dynamical state of the cluster, we analyse \textit{XMM-Newton} data, \textit{Chandra} data, and Sloan Digital Sky Survey data. Moreover, we compare our results to clusters extracted from the {\sc The Three Hundred Project} cosmological simulation. 
     }{
     % Results
     We find that Abell\,1430 consists of two components, namely A1430-A and A1430-B, with a mass ratio of about 2:1. The massive component shows diffuse radio emission which can be classified as radio halo which shows a low radio power at 1.4\,GHz with respect to the mass of the cluster. Most interestingly, there is  extended diffuse radio emission, in the following dubbed as the `Pillow' according to its morphology, which is apparently related to A1430-B and which is neither typical halo nor typical relic emission. The origin of this emission is puzzling. We speculate that the two components of Abell\,1430 undergo an off-axis merger. In this scenario, A1430-B is moving towards the main cluster component and may have compressed and stirred the medium in the filament between the two cluster components.
     }{
     % Conclusions 
     We have discovered evidence for diffuse radio emission related to the low-density intracluster or intergalactic medium in Abell\,1430. To date, only few examples of emission originating from such regions are known. These discoveries are crucial to constrain possible acceleration mechanisms which may allow us to explain the presence of relativistic electrons in these regions. In particular, our results indicate a spectral index of $\alpha_{144\,\text{MHz}}^{1.5\,\text{GHz}}=-1.4\pm0.5$ for the Pillow. If upcoming observations confirm a slope as flat as $-1.4$ or even flatter, this would pose a challenge for the electron acceleration scenarios.  
     }
    
   % 5 {} token are mandatory


   \maketitle
%
%-------------------------------------------------------------------

\section{Introduction}

  % galaxy cluster surveys 
  Galaxy clusters have been studied extensively thanks to the rigorous surveys that have observed the entire sky to search for overdense regions \citep{abell_1958, noras_2000, ebeling_2000, 2011A&A...527C...2Z, planck_2011, 2017AJ....153..220B} and many follow-up deep studies of individual clusters.   The analysis of the thermal gas, filling the volume between the cluster galaxies, permits to conclude on the dynamical state of the clusters and the effects that occur in the course of merging \citep[see, e.g.,][]{2002ApJ...567L..27M,2019ApJ...882...69G}. 
  
  % many merging clusters show halo and relics 
  For many merging clusters diffuse radio emission has been found.   The emission features can generally be divided into radio relics and radio halos, which are attributed to merger shock fronts in the cluster periphery and to volume-filling turbulence in the intracluster medium (ICM), respectively \citep[see, e.g.,][for recent reviews]{feretti_2012,vanWeeren2019}. Both relics and halos show spectral properties which indicate that synchrotron emission is the origin, hence, they manifest the presence of magnetic fields and relativistic electrons in the ICM. 
  
  % radio relics, properties: assoc. w. shocks, polarised 
  For several radio relics, a connection to a shock front in the ICM is evident from the X-ray surface brightness or temperature discontinuities at the location of the relic  \citep[e.g.,][]{2010ApJ...715.1143F,2013ApJ...764...82B,2013PASJ...65...16A,2016MNRAS.463.1534B}.   The typically very elongated morphology of relics is hence naturally explained by relating the relic to a spherical shock observed in projection.   A radio emitting shock seen face-on is expected to have such a low surface brightness that is difficult to detect.  Radio relics show in general power-law spectra with a slope of about $-1.0$ to $-1.3$ \citep{vanWeeren2019}.   A very prominent feature of radio relics is that often the emission has found to be polarised in the GHz-regime, for some relics locally even with a fractional polarisation above 50\,\% \citep{2009A&A...494..429B,2010Sci...330..347V,2012A&A...546A.124V,2017A&A...600A..18K,2018MNRAS.478.2218H}.   In several cases, the polarisation angle orientation is remarkably homogeneous across the entire relic \citep[e.g.][]{2010Sci...330..347V}. 
  
  % radio halos, properties
  In contrast, radio halos are believed to trace the turbulence generated by cluster mergers.   A direct observational proof is still to be provided.   However, radio halos have almost exclusively been found in merger systems \citep{buote_2001,2010ApJ...721L..82C}, suggesting that the halo emission is related to the kinetic energy dissipated during the merger and transferred into non-thermal components through complex mechanisms involving shocks and turbulence \citep[see][for a review]{brunetti_2014}.   This scenario can explain the observational properties of halos, e.g., the often found close correlation between X-ray and radio surface brightness and the absence of polarised halo emission at the current detection levels.   Moreover, it has been found that the luminosity of radio halos correlates with the X-ray luminosity and the mass of the clusters \citep{2013ApJ...777..141C}. 
  
  % diffuse emission in low density region, radio bridges
  Recently, LOw Frequency ARray (LOFAR) observations also discovered diffuse emission which connects pairs of massive clusters in the form of radio bridges \citep{botteon_2018,2019Sci...364..981G,2020MNRAS.499L..11B}.  These observations suggest that compression and turbulence generated by substructures in the low-density cosmic filaments activate magnetic field amplification and mechanisms of stochastic acceleration of particles, too  \citep{2020PhRvL.124e1101B}. To date, it is unclear 
  how often such early stages of cluster mergers lead to observable radio emission that originates from turbulence driven by the motion of the substructures.   Since this emission is expected to possess a steep spectrum \citep{2020PhRvL.124e1101B}, LOFAR observations are excellently suited to search for new examples.

  In this paper, we present the discovery of a radio halo in the merging galaxy cluster Abell\,1430. Central for this study are observations being part of the LOFAR Two-Metre Sky Survey (LoTSS). We complemented our analysis with data from Karl G. Jansky Very Large Array (VLA) observations in L-band. \textit{Chandra} and \textit{XMM-Newton} X-ray data was also used, allowing us to examine the relationship between the thermal properties of the cluster components with the relativistic electron population.
  
  Throughout this paper, we assume a $\mathrm{\Lambda CDM}$ cosmology with $H_{0}=70\,\mathrm{km\,s^{-1}\,Mpc^{-1}}$, $\Omega_{m}=0.3$, and $\Omega_{\Lambda}=0.7$. With these values, $1\arcsec$ corresponds to a physical scale of 5\,kpc at a redshift of 0.35 \citep{2006PASP..118.1711W}. We apply the usual convention for the spectral index $\alpha$ of a power-law spectrum, namely $S_{\nu} \propto \nu^{\alpha}$. All images shown in this paper are in the J2000 coordinate system. Radio images are corrected for primary beam attenuation.

% ===========================================================
%
%  A1430
%
% ===========================================================

\section{Abell 1430}
\label{sec::abell_1430}

  % A1430 detection: Abell, ROSAT
  Abell\,1430, hereafter A1430, was first reported in the catalogue of 2712 rich clusters of galaxies found in the National Geographic Society Palomar Observatory Sky Survey \citep{abell_1958}. The catalogue lists 31 cluster galaxies making A1430 have a richness class of 0, i.e., the number of galaxies is actually below the minimum population of 50 galaxies to be considered as a rich cluster. The ROSAT space telescope has detected X-ray emission from the cluster; in the Northern ROSAT All-Sky galaxy cluster survey (NORAS), it is listed as an extended X-ray source, namely RXC\,J1159.2+494 \citep{noras_2000}. 
  
  % A1430 part of Planck cluster catalogue
  The cluster has been also reported in the First and Second Planck Catalogue of Sunyaev-Zel'dovich effect (SZ) sources, PSZ1 and PSZ2, respectively \citep{2015A&A...581A..14P,2016A&A...594A..27P}. In the second catalogue, it is listed as PSZ2\,G143.26+65.24 with a redshift of $z_{\rm PSZ2}=0.363$ and a mass of $M_{\rm SZ,500}=(7.6\pm0.4)\times 10^{14} \,\textup{M}_\odot$. 

  % earlier a lower redshift
  In earlier analyses, a lower redshift was derived.  \citet{1991ApJS...77..363S} compiled redshifts and velocity dispersions of Abell clusters. They stated a redshift of $0.21$ for A1430, derived from the photometric redshift of two galaxies, which later turned out to be located in the foreground of A1430.  \citet{rozo_2015} compared clusters from the PSZ1 catalogue with their Sloan Digital Sky Survey (SDSS) redMaPPer catalogue. For twelve clusters, they found a discrepancy between the redshift given in PSZ1 and their results; A1430 was among these clusters. Based on the spectroscopic redshift of galaxies, the redshift of A1430 was recalculated as $z_{\rm Rozo}= 0.350\pm 0.014$. 
  
  % BCG Xray peak shift, Rossetti+ 16
  \citet{2016MNRAS.457.4515R} investigated the projected distance between the positions of the brightest cluster galaxy (BCG) and the peak of the X-ray surface brightness for 132 galaxy clusters.  For A1430 they found a distance of $0.034\,R_{500}$, which is significantly larger than the median $0.017 \, R_{500}$ of all clusters in their sample and larger than the threshold of $0.02 \, R_{500}$ given by \citet{2009MNRAS.398.1698S} for separating relaxed and disturbed clusters. 
  
  % wrap up Sec: need to know merger state to understand origin to conclude on origin of radio emission 
  The shift between the BCG and the X-ray emission peak indicates that the cluster is dynamically disturbed, however, it does not allow us to conclude on details of the merger state. We present in Sect.\,\ref{sec::xrayobs} and Sect.\,\ref{sec::redshift_separation}, a detailed study of the X-ray emission and of the redshift distribution to determine the actual dynamical state of A1430.
  
  % LoTSS
  A1430 was observed as part of the LOFAR Two-metre Sky Survey (LoTSS). The cluster lies in the area which has been covered by the first LoTSS data release \citep{2019A&A...622A...1S}.

% ==============================================================
%
%  Observations, data reduction an resulting images 
%
% ============================================================== 

\section{Observations, data reduction and resulting images}
\label{sec::observations_data_reduction}

% ======= SUBSEC :: LOFAR  =====================================
%
\subsection{LOFAR High Band Antenna (HBA) observations}
\label{sec::lofarobs}

  % ======= FIGURE :: LoTSS-highres  ===========================
  \begin{figure*}[t]
    \centering
    \includegraphics[width=0.8\textwidth]{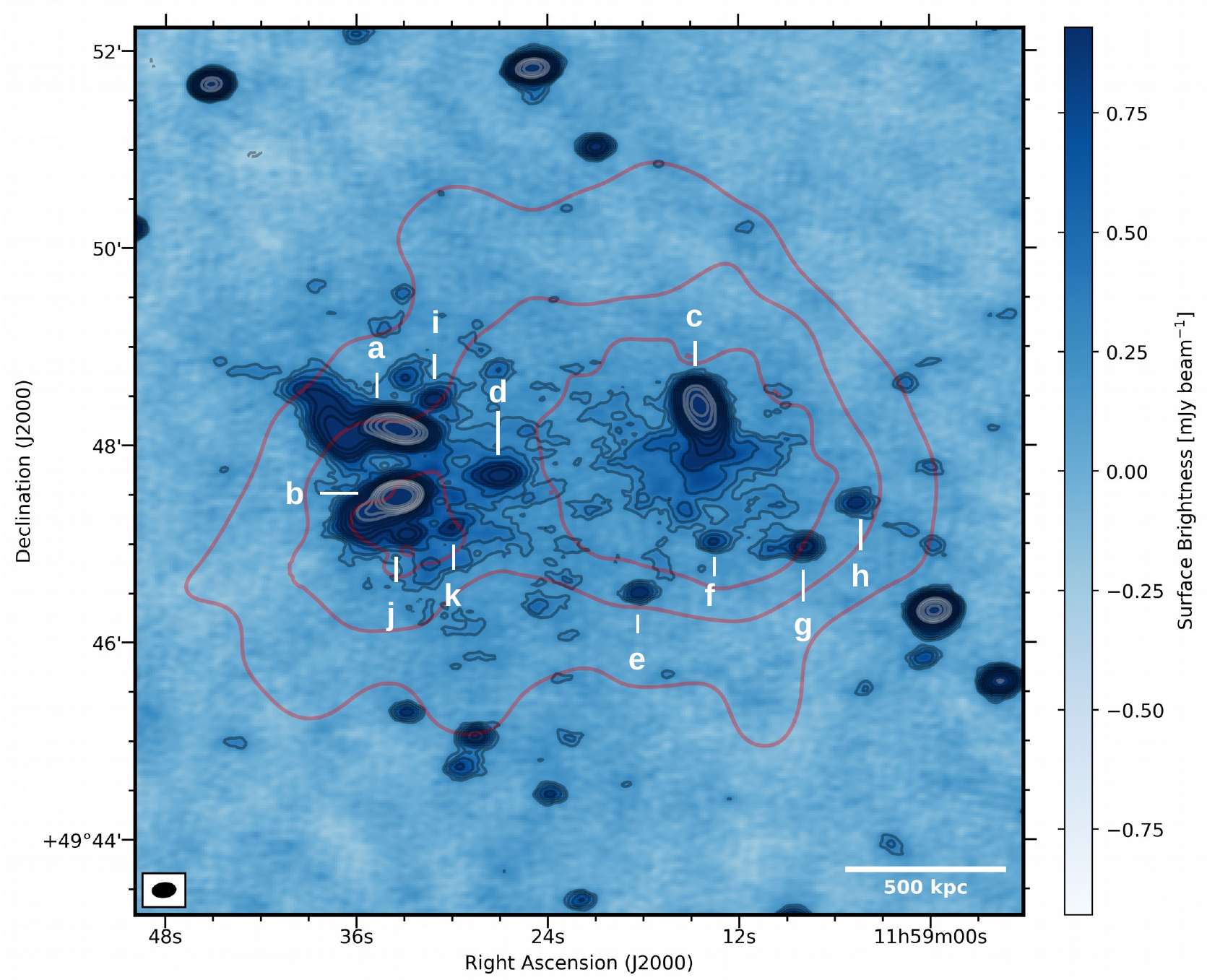}
    \caption{
      Total intensity radio map of A1430 at a central frequency of 144\,MHz as obtained from the LoTSS survey. There are three bright, tailed radio galaxies, labelled as a, b and c, in the cluster region and several additional compact sources (brighter ones are marked with d-h). Moreover, there is evidence for diffuse radio emission in the cluster region. The image is obtained using a {\tt Briggs} weighting of $0{.}0$.  The noise level is $\sigma_{\rm{rms}}=$\,92\,\textmu J\,beam$^{-1}$ and the restoring beam has a size of $14\arcsec \times 9\arcsec$ and a position angle of $-84^{\circ}$, as illustrated by the black ellipse in the bottom left corner of the image. Black contours start from $3\sigma_{\rm{rms}}$ and are spaced by $\sqrt{2}$. Contours at higher levels are drawn in grey for an optimal visual presentation. No negative contour levels below $-3\sigma_{\rm{rms}}$ are present.  Red contours represent the distribution of X-ray emission as seen by \textit{Chandra} to indicate the cluster region. Flux densities of the annotated sources are listed in Table\,\ref{tab::compactsources}.
      }
    \label{fig::LoTSS-highres}
  \end{figure*}

  % LoTSS 
  LoTSS is mapping the entire northern sky with unprecedented sensitivity and resolution. At optimal declinations, it provides images with root mean square (rms) noise levels below 100\,\textmu Jy at a resolution of 6$\arcsec$ and 20$\arcsec$ at a central frequency of 144\,MHz \citep{2019A&A...622A...1S}. In order to cover the entire northern sky, 3,168 pointings of eight hours each have to be observed. Every LoTSS pointing is bookended with calibrator observations of 10 minutes each. LoTSS data for 2\,\% of the northern sky, covering about 424 square degrees in the region of the Hobby-Eberly Telescope Dark Energy Experiment (HETDEX) Spring Field has been published as LoTSS Data Release 1 \citep[DR1, ][]{2019A&A...622A...1S}.
  
  % LoTSS A1430 observations, part of DR1, removed P19Hetdex17, P21,
  The field of A1430 was covered by two LoTSS pointings located in the HETDEX field, namely P18Hetdex03, and P22Hetdex04. Those observations were conducted on 28th of May 2014 and 25th of May 2014, respectively. 
  
  % LOFAR/LoTSS data reduction 
  Data reduction and calibration was performed with the LoTSS DR2 reduction pipeline, which contains significant improvements in calibration and imaging reconstruction fidelity \citep{2021A&A...648A...1T} as compared to the LoTSS DR1 \citep[see Sec. 2.2, 2.3, and 5 in][]{2019A&A...622A...1S}. The calibration comprises a direction-independent calibration using \texttt{PREFACTOR}\footnote{https://github.com/lofar-astron/prefactor/} v2.0 \citep[for a description of the procedure, see][]{reinout_2016, williams_2016} and the \texttt{DDF-pipeline}\footnote{https://github.com/mhardcastle/ddf-pipeline/} v2.2, developed by the LOFAR Surveys Key Science Project, which performs several iterations of direction-dependent self-calibration, and imaging using \texttt{KillMS} \citep{2014arXiv1410.8706T,2014A&A...566A.127T,2015MNRAS.449.2668S} and \texttt{DDFacet} \citep{2018A&A...611A..87T}.

  % extraction and imaging 
  The two directional-dependent calibrated data sets were combined and an additionally common calibration \citep[see Section 2 in][for a detailed description of the ``extraction'' method]{2020arXiv201102387V} was performed to further improve the image fidelity in the target region.   As part of this pipeline, a source model of the whole field of view, as derived from the \texttt{DDF-pipeline}, and excluding an area of 0.5\,deg$^{2}$ centred on the location of A1430, is subtracted from the visibilities for each of the two sets. Afterwards, the visibilities of all pointings are phase-shifted and averaged towards the target direction. Subsequently, several rounds of imaging and self-calibration with respect to amplitude and phase were performed using the combined data of both pointings. Final imaging was performed with \texttt{WSClean v2.6} \citep{offringa-wsclean-2014} using different Briggs weightings \citep{1995AAS...18711202B} for high-resolution and low-resolution imaging. As described below in more detail, we carefully modelled the tailed radio galaxies and compact sources and subtracted them from the visibilities with \texttt{DPPP}\footnote{https://github.com/lofar-astron/DP3} \citep{2018ascl.soft04003V} before imaging at low resolution. 
  
  % LOFAR clusters in HETDEX paper
  \citet{2020arXiv201102387V} have characterised the diffuse radio emission revealed in the LoTSS observations for all known galaxy clusters in the HETDEX region, evidently including A1430. In that work, the procedures for extracting the information for the region of interest from the LoTSS data and applying self-calibration in addition to the standard LoTSS pipeline are described in detail. For the analysis of the radio emission in A1430, i.e., the focus of this work, we start with the extracted and self-calibrated data.
  
  % LoTSS image  
  %   figure LoTSS-highres
  %   introduce compact sources, 
  %   just mention diffuse em in merging region
  %   highlight difficulty for disentangeling compact and diffuse emission
  The radio emission at 144\,MHz in the A1430 region -without the subtraction of compact sources- is shown in Fig.\,\ref{fig::LoTSS-highres}. Most prominently, three bright radio sources are located in the cluster region, denoted as a, b, and c.   All three sources are evidently extended and appear to be tailed radio galaxies.  The low surface brightness, extended features may indicate diffuse emission in the cluster region not related to the bright radio galaxies.   A cautious modelling, in particular of the tails of the radio galaxies, is crucial since the bright radio galaxies overlap with the diffuse emission.  The low surface brightness diffuse emission can be recovered in maps made at lower resolution only if the bright radio galaxies and other compact sources in the cluster region can be reasonably well subtracted from the visibilities.

  In Tab.\,\ref{tab::compactsources} we list the compact and extended sources in the cluster region identified from the LoTSS image and subtracted from the visibilities. The flux density uncertainties are estimated as follows:
  \begin{equation}
    ( \Delta S )^2 
    = 
       \left(f_{\text{scale}} \cdot S\right)^{2} + 
       \sigma_{\text{rms}}^2 \cdot N_{\text{b}} +
       \left(f_{\text{sub}} \cdot S_{\text{sub}}\right)^{2} \:\:
     ,
    \label{eq::fluxerror}
  \end{equation}
  where $S$ denotes the flux density of the source, $f_{\text{scale}}$ the flux density scale uncertainty, $\sigma_{\text{rms}}$ the noise in the image, $N_{\text{b}}$ the number of beams necessary to cover the source, $f_{\text{sub}}$ the subtraction uncertainty if applicable, and $S_{\text{sub}}$ the subtracted flux density if applicable. For the flux densities derived from the LOFAR data, we assume a flux density scale uncertainty of $20\,\%$ as done by LoTSS \citep{2019A&A...622A...1S} and a subtraction uncertainty of $10\,\%$.

  % ======= TABLE: compact sources  ==============================
  \setlength{\tabcolsep}{7pt}
  \begin{table}[bp]
    \caption{
      Flux densities of the sources marked in Fig.\,\ref{fig::LoTSS-highres} for both LOFAR HBA observation at 144\,MHz as well as for the VLA at 1.5\,GHz. % \textcolor{red}{complement with redshifts from SDSS in case known (Matthias)} 
      }
    \centering
    \begin{tabular}[]{c|D{X}{\,\pm\,}{-1}|D{X}{\,\pm\,}{-1}}
    \hline \hline
    name & \multicolumn{1}{c|}{$S_{\rm{144\,MHz}}$} & \multicolumn{1}{c}{$S_{\rm{1.5\,GHz}}$}  \\
     & \multicolumn{1}{c|}{ [mJy] }  & \multicolumn{1}{c}{ [mJy] }    \\[.3ex]
    \hline
    a   & 66.1X          13.3 &  8.6\phantom{0}X0.6      \\ 
    b   & 89.7X          18.0 & 15.4\phantom{0}X2.0      \\
    c   & 38.6X\phantom{0}7.8 &  9.6\phantom{0}X1.1      \\
    d   &  3.8X\phantom{0}0.8 &  0.2\phantom{0}X0.1    \\
    e   &  1.2X\phantom{0}0.3 &  0.1\phantom{0}X0.1    \\
    f   &  0.8X\phantom{0}0.2 &  0.3\phantom{0}X0.1    \\
    g   &  2.0X\phantom{0}0.5 &  0.4\phantom{0}X0.1    \\
    h   &  1.2X\phantom{0}0.3 &  0.2\phantom{0}X0.1    \\
    i   &  1.3X\phantom{0}0.3 &  0.2\phantom{0}X0.1    \\
    j   &  2.2X\phantom{0}0.5 &  0.3\phantom{0}X0.1    \\
    k   &  1.1X\phantom{0}0.3 &  0.1\phantom{0}X0.1    \\
    \hline \hline
    \end{tabular}
    \label{tab::compactsources}
  \end{table}

\subsection{Uncovering the diffuse emission}
\label{sec::uncoverdiffuse}

% ======= FIGURE: Subtracted Model  ======================================= 
  \begin{figure}[htbp]
    \centering
    \includegraphics[width=0.48\textwidth]{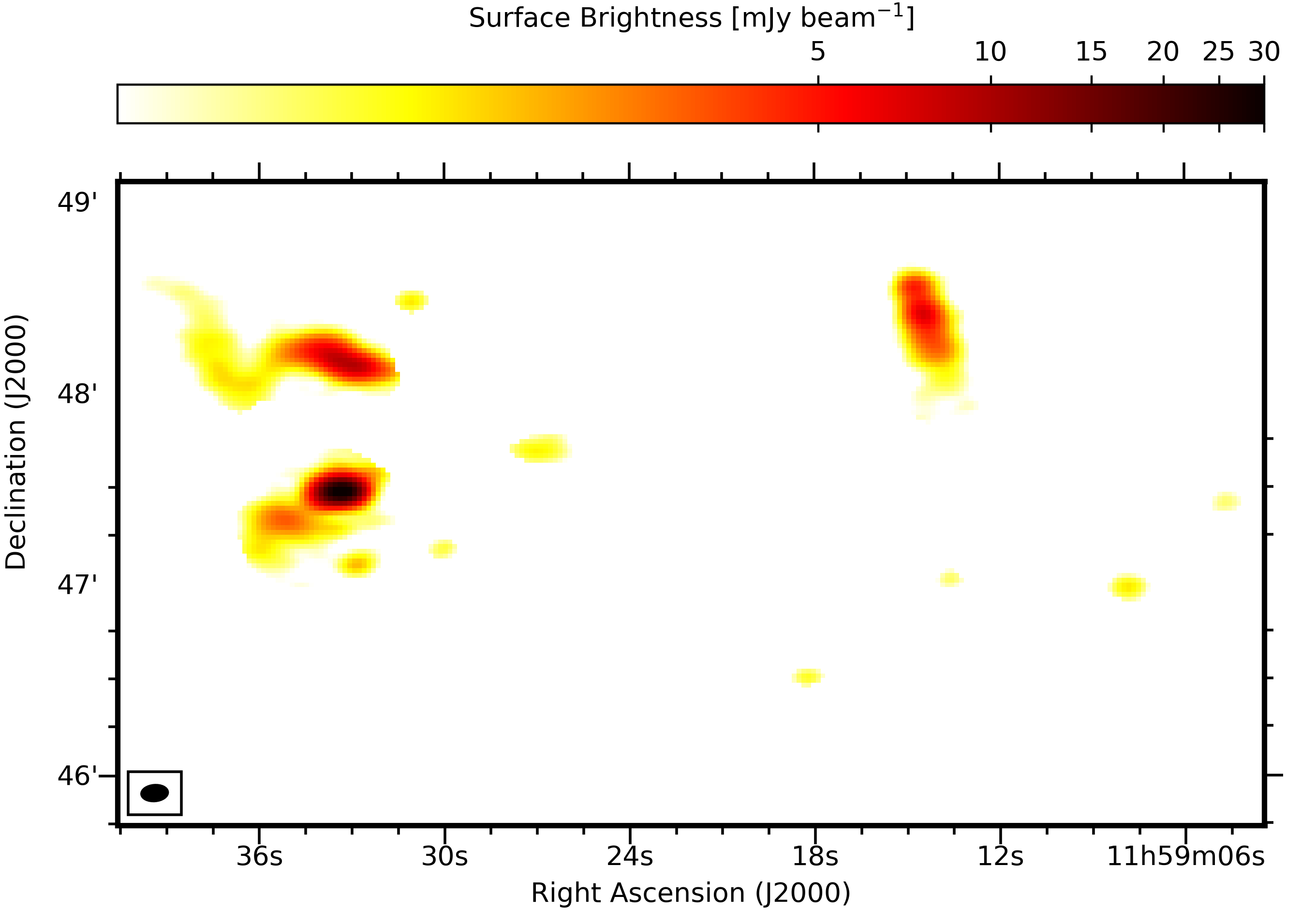}
    \caption{Surface brightness distribution of the compact sources and radio galaxies marked in Fig.\,\ref{fig::LoTSS-highres}, which was subtracted from the $uv$-data for subsequent low-resolution imaging. Here, the model components were convolved with a restoring beam of $8^{\prime\prime}\times5^{\prime\prime}$ for illustration purposes.
    }
    \label{fig::subtractmodel}
  \end{figure}

 The radio surface brightness distribution shown in Fig.\,\ref{fig::LoTSS-highres} tentatively indicates the presence of extended, diffuse emission in the cluster. However, it is challenging to distinguish diffuse emission in the cluster from extended lobes of the galaxies. To achieve this, we (i) carefully model the sources, (ii) compare our subtraction to the $uv$-cut based subtraction applied in \citet{2020arXiv201102387V}, and (iii) discuss the plausibility that the detected extended emission could actually originate from the lobes of the radio galaxies. 
 
 A common procedure to distinguish emission of radio galaxies from extended, diffuse emission is to apply an inner cut in $uv$-space, i.e., all visibilities on $uv$-distances larger than the cut value are used to generate the surface brightness distribution of the radio galaxies while all visibilities on shorter distances are used to reconstruct the extended, diffuse emission. In \citet{2020arXiv201102387V} we adopted a $uv$-cut corresponding to 250\,kpc at the cluster redshift. For instance, we identified for the radio galaxy c a flux density of 32.6\,mJy based on this method. Using a larger $uv$-cut of 500\,kpc leads to a similar flux density for source c. However, the $uv$-cut method may underestimate the flux density related to source c, because part of the flux density at the location of source c is attributing to an `extension' of the diffuse emission. Due to emission of source c not subtracted with the model the diffuse emission may seem to be more extended than it actually is. 
 
 It is impossible to decide from the radio map if the radio galaxy is superimposed with the halo or not. We adopted a rigorous approach to subtract the radio galaxies by basically model all flux in the galaxy region, identified from the $uv$-cut, as part of the radio galaxies, see Fig.\,\ref{fig::subtractmodel}. With this approach we obtained the flux densities listed in Tab.\,\ref{tab::compactsources}. Apparently, we subtracted for source c about 6\,mJy more than obtained with the $uv$-cut method.
 
 The difference between the two methods is even more severe for the sources a and b. With the two $uv$-cuts we obtained a flux density of 130\,mJy and 137\,mJy for both sources together, using an $uv$-cut of 250\,kpc and 500\,kpc, respectively. With our more conservative approach, we subtracted 155.8\,mJy. As mentioned above, it is impossible to unambiguously determine how much of the flux density in the region of the galaxies actually originates from the diffuse emission. We therefore decided to conservatively attribute all flux density in the region of the galaxies to the galaxies. 

 Evidently, the question remains if the diffuse emission might also be partially -or even fully-  related to the radio galaxies, e.g., being an old extended lobe. We will discuss below the plausibility of such a scenario.

% ======= SUBSEC: VLA  ===========================================
% 
\subsection{VLA observations}
\label{sec::vlaobs}

  % ======= FIGURE: VLA medium res with compact sources  =====
  \begin{figure}[t]
    \centering
    \includegraphics[width=0.48\textwidth]{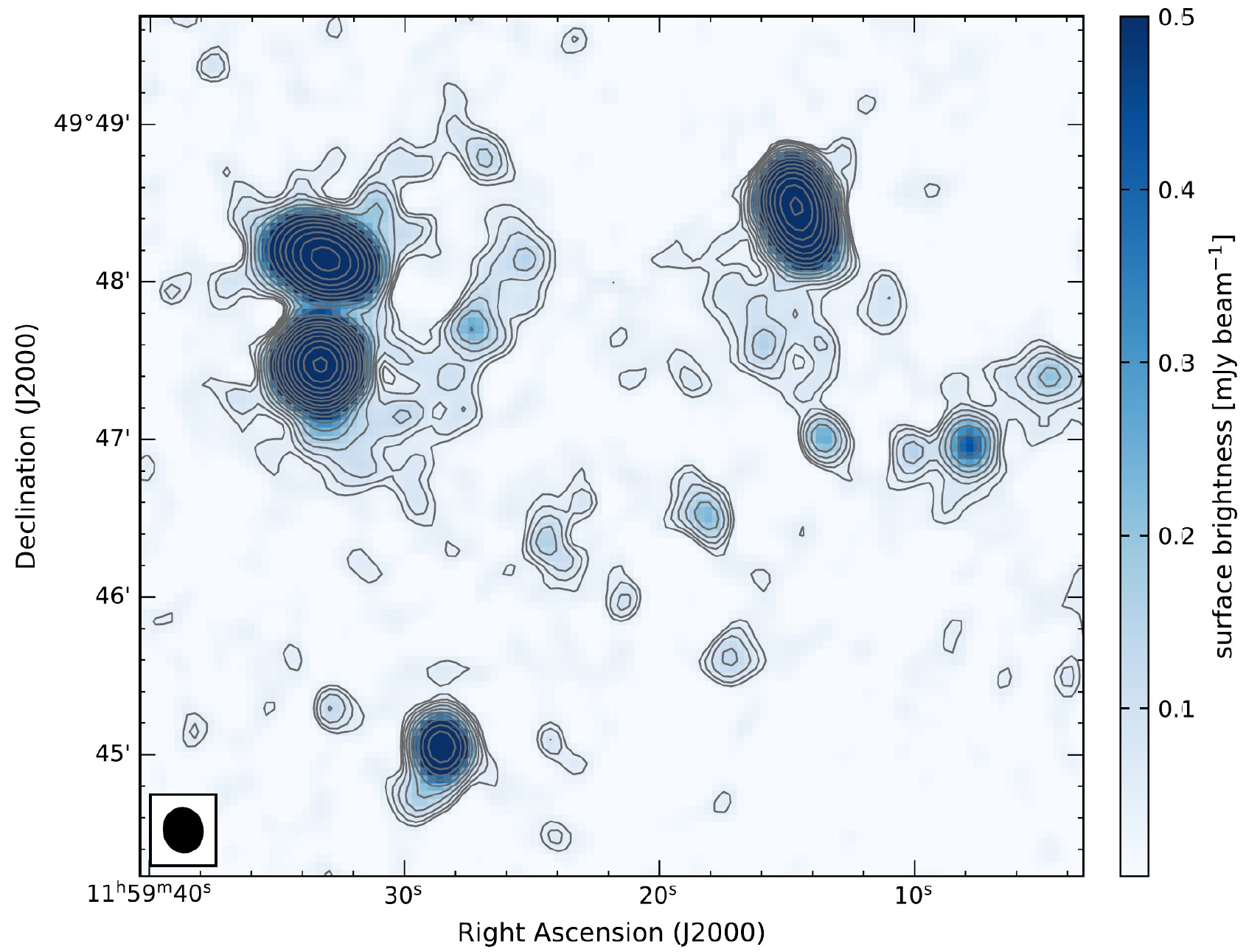}
    \vspace{-0.2cm}
    \caption{VLA L-band, C and D configuration image of the cluster A1430. The bright radio galaxies but and evidence for diffuse emission are visible. The flux densities of the compact sources, as labelled in Fig.\,\ref{fig::LoTSS-highres}, are listed in Tab.\,\ref{tab::compactsources}. The restoring beam size in the image is $ 16\arcsec \times 14\arcsec $ and the noise level is $\sigma_{\text{rms}}=16$\,\textmu Jy\,beam$^{-1}$. The image is created using {\tt robust}\,=\,0. Contour levels start at $3\, \sigma_{\rm{rms}}$ and are spaced by a factor of $\sqrt{2}$. No negative contour levels below $-3\sigma_{\rm{rms}}$ are present.
      }
    \label{fig::VLAimagea}
  \end{figure} 
 
  We observed the cluster with the VLA in the L-band covering the frequency range of 1-2\,GHz. These observations (project code: 18A-172) were carried out with C and D configurations on September 2018 and January 2019, respectively, with in total 2.6\,h observing time on target. All four correlation products, namely RR, RL, LR, and LL, were recorded.   The data were recorded with 16 spectral windows, each divided into 64 channels. For each configuration, 3C147 was observed as the primary calibrator. 3C286 and J1219+4829 were included as secondary calibrators.  

  % data reduction
  The VLA data were reduced with $\tt{CASA}$ version 5.1.0.   The data from C- and D-configurations were calibrated independently but in the same way.   We first determined and applied elevation-dependent gain tables and antenna offsets positions.   The data were then inspected for radio frequency interference (RFI) removal.   The software $\tt{AOFlagger}$ \citep{Offringa2010} was used for accurately detecting and flagging of RFI.   We used the L-band 3C147 model provided by the $\tt{CASA}$ software package and set the flux density scale according to \citet{Perley2013}.   We calibrated the initial phase, parallel-hand delays and bandpass using 3C147.   Gain calibration was obtained for 3C147 and secondary calibrators.   For polarisation calibration, the leakage response was determined using the unpolarised calibrator 3C147. The cross-hand delays and the absolute position angle were corrected using 3C286.   The calibrated solutions were then transferred to the target field. 
  
  % imaging 
  Following initial imaging, we performed two rounds of phase-only self-calibration and two of amplitude and phase self-calibration on the individual data sets.   The imaging was performed using the $\tt{Briggs}$ weighting with $\tt{robust}=0.0$, $\tt{wprojplanes}=250$, and $\tt{nterms}=3$, see Fig.\,\ref{fig::VLAimagea} for the resulting image.  To recover the diffuse emission, we combined the C and D-configurations data and imaged in $\tt{WSClean}$ using $\tt{robust}=0.6$ (similar to the LOFAR image).   We assume a flux density scale uncertainty of $4\,\%$ for the VLA L-band \citep{Perley2013}.

% ======= SUBSEC: XMM and Chandra  ===============================
%
\subsection{\textit{XMM-Newton} and \textit{Chandra} observations}
\label{sec::xrayobs}

  % ======= FIGURE: XMM  ========================================= 
  \begin{figure*}[htbp]
    \centering
    \includegraphics[width=0.48\textwidth]{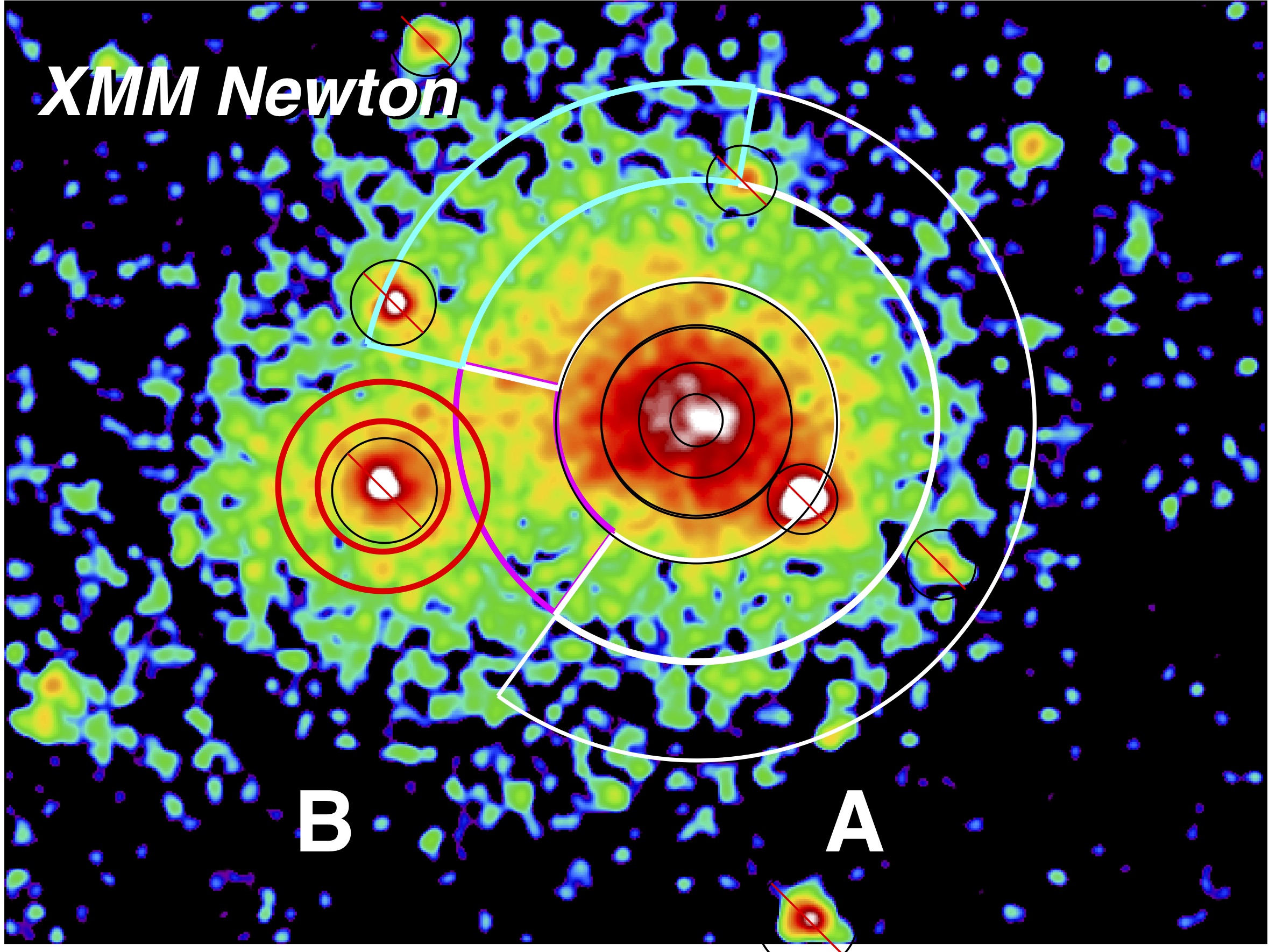}
    \includegraphics[width=0.48\textwidth]{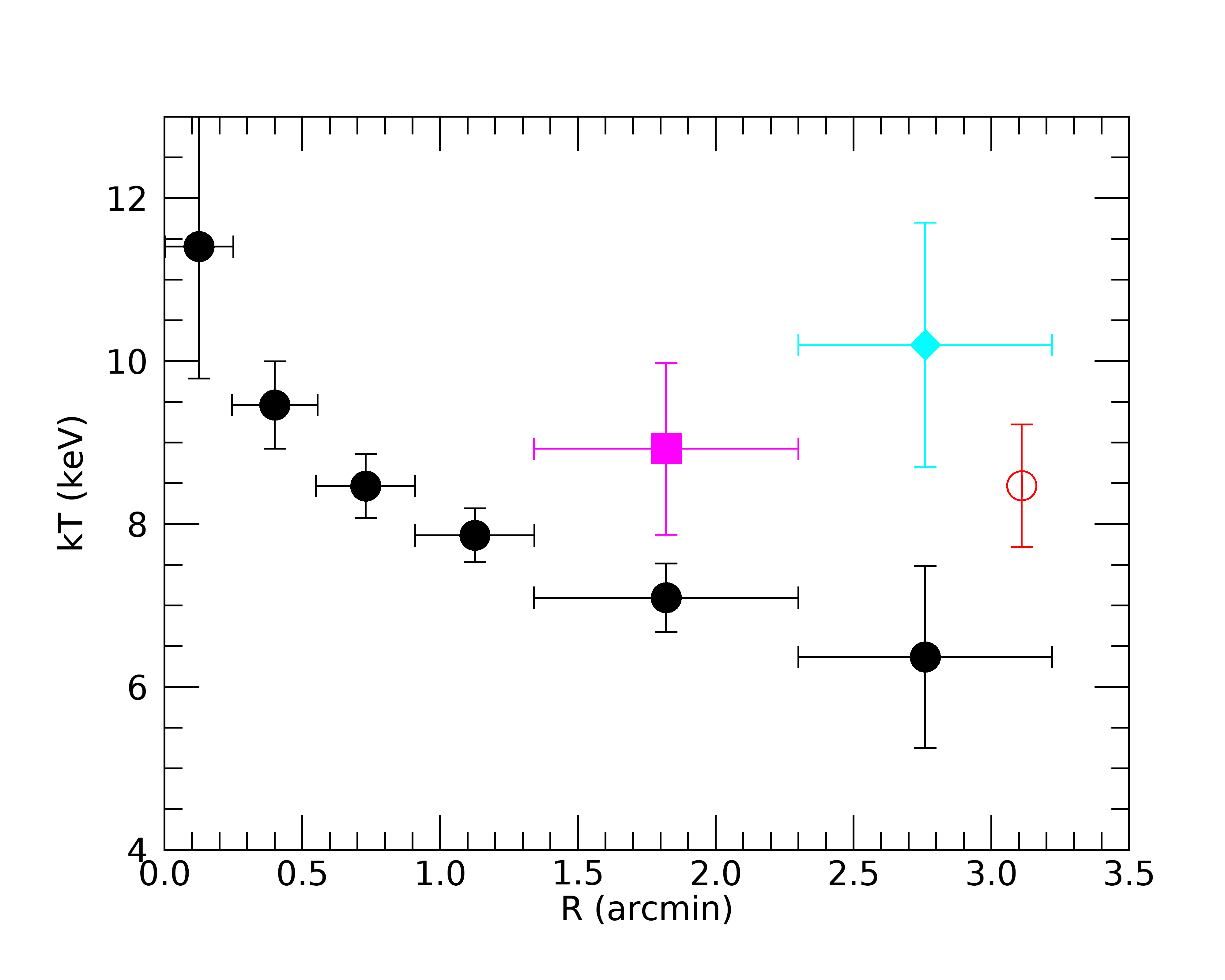}
    \caption{
      {\it Left:} X-ray surface distribution as obtained from \textit{XMM-Newton} in the band $0.7-1.2$\,keV, smoothed with a Gaussian of a width of $12\arcsec$.   The two subcomponents of A1430 are clearly visible. Annuli and sectors show the regions used for spectral extraction and analysis, while barred circles indicate the point sources excluded from the analysis.   The labels `A' and `B' denote the main component and the subcomponent of the cluster, respectively. {\it Right:} Temperature distribution in the cluster.   Filled black circles show the temperature profile of the main cluster, measured in the black and white marked regions in the left panel which exclude the interaction with the subcomponent.   The magenta and cyan regions show the temperature in the sectors marked with the same colours in the left panel, while the red empty circle represents the temperature of the subcomponent A.}
    \label{fig::XMM}
  \end{figure*}
  
  % XMM Heritage, observation and data reduction 
  A1430 has been observed as part of the XMM Heritage Cluster Project\footnote{http://xmm-heritage.oas.inaf.it}, a large and unbiased sample of 118 clusters, detected with a high signal-to-noise ratio in the Second Planck SZ Catalogue.   The observation was split into two OBSIDs ($0827320201$ and $0827020201$), with a total clean exposure time of $54.6\,\rm{ks}$ with MOS1, $84.5\,\rm{ks}$ with MOS2 and $24.4\,\rm{ks}$ with pn\footnote{The pn and MOS1 camera were switched off for technical reasons for part of the observations}.   We analysed the data with Science Analysis System (SAS) version $16.1$, following the procedures in \citet{ghirardini_2019} for data reduction, image production, and spectral extraction and fitting.   These procedures were designed for the mosaic observations of the XMM Cluster Outskirts Project (X-COP) project \citep{eckert_2017}: they combine the spectra for different observations by jointly fitting them and produce mosaic images.
 
  % XMM image 
  In the left panel of Fig.\,\ref{fig::XMM}, we show the mosaic image in the X-COP band $0.7-1.2\,\rm keV$, which maximises the source-to-background ratio \citep{ghirardini_2019}.   We run the source detection algorithm on this image with the SAS tool \verb|ewavelet| finding a few candidate sources embedded in the ICM.   Given the moderate spatial resolution of XMM, it is hard to assess if they are bona fide point sources or rather extended features, possibly part of the ICM of A1430.   We thus checked all of them with the higher resolution \textit{Chandra} image, see below, and masked all sources flagged in the left panel of Fig.~\textcolor{red}{\ref{fig::XMM}} in the subsequent analysis. 
  
  % ======= FIGURE: Chandra  =================================
  \begin{figure}[tbp]
    \centering
    \includegraphics[width=0.48\textwidth]{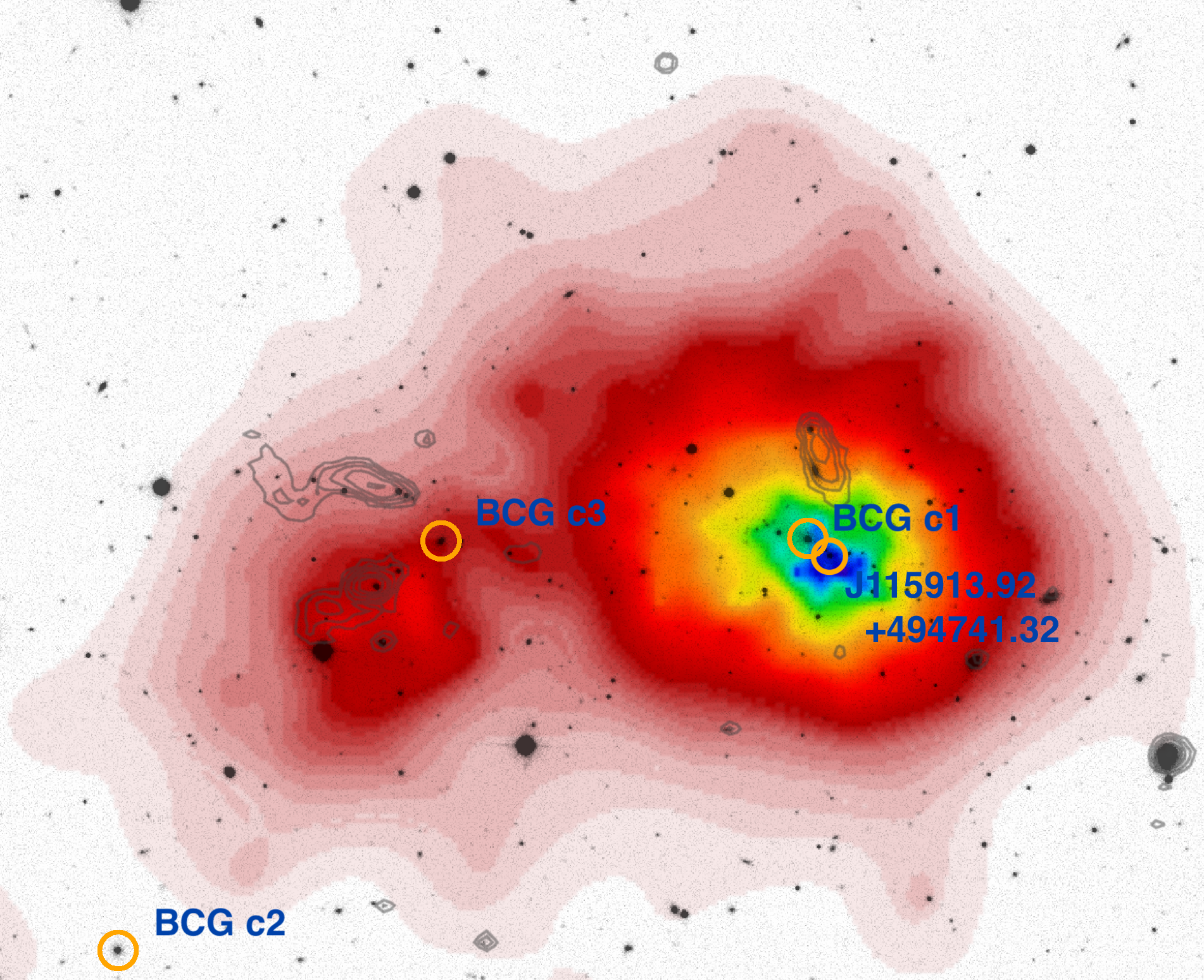}
    \caption{
      {\it Colorscale:} \textit{Chandra} X-ray surface brightness distribution with point sources subtracted and adaptively smoothed.    {\it Grey contours:} LOFAR high-resolution image (using Briggs weighting $-1.0$) to indicate the positions of the radio galaxies. {\it Greyscale} SDSS $r$-band image. The positions of the three initial BCG candidates c1, c2, and c3 at J115914.85+494748.1, J115943.99+494459.6, and  J115930.36+494747.4, respectively, and the bright galaxy close to the X-ray peak are highlighted with orange circles, see Sec.\,\ref{sec::redshift_separation}.
      }
    \label{fig::Chandra}
  \end{figure}
  
  % Chandra 
  A1430 was observed with the \textit{Chandra} X-ray Observatory in January 2014 (Obs-ID 15119) for 22\,ks. The data reduction has been performed as described in \citet{2005ApJ...628..655V}.   We applied the calibration files of CALDB\,4.7.2. The data reduction included corrections for the time dependence of the charge transfer inefficiency and gain, and also a check for periods of high background, which were then removed (only $45\,\rm s$ were discarded). Standard blank sky background files and read-out artifacts were subtracted. Fig.\,\ref{fig::Chandra} shows the resulting, adaptively smoothed X-ray surface brightness distribution after the subtraction of point sources, using \texttt{dmimgadapt}.

% ======= SUBSEC: SDSS  ================================== 
%
\subsection{SDSS}
\label{sec::sdss}

  % ======= FIGURE: sdss_redshifts  ======================
  \begin{figure}[t]
    \centering
    \includegraphics[width=0.48\textwidth]{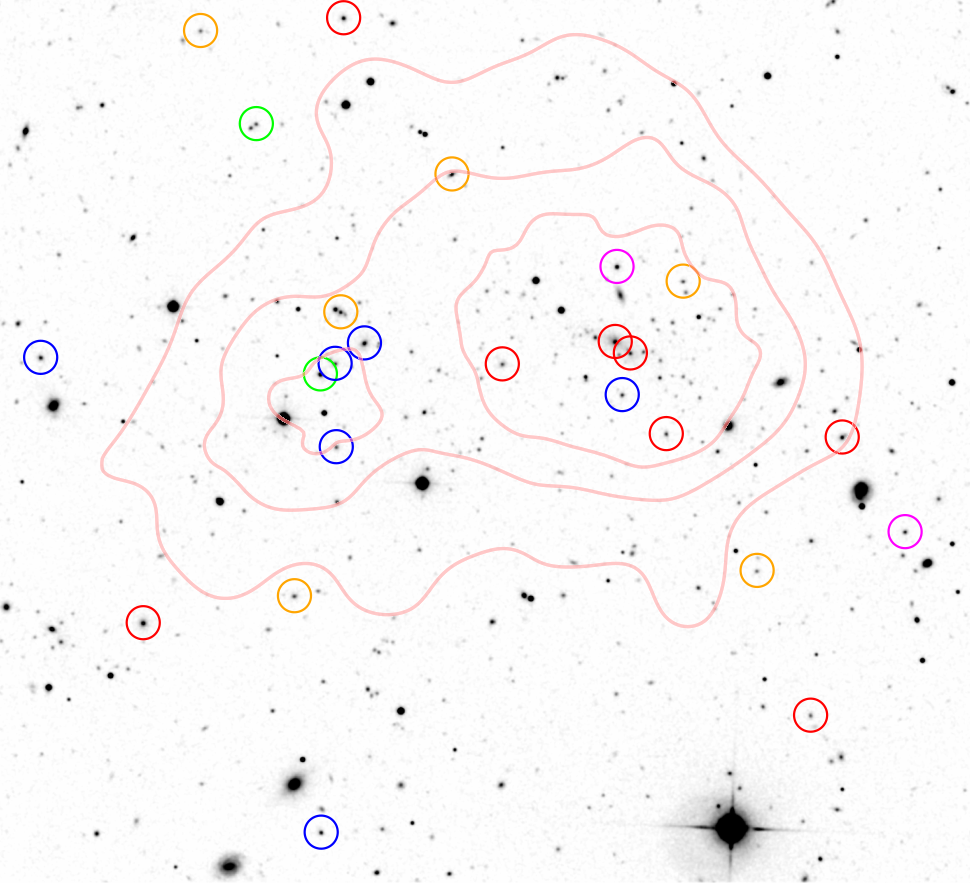}
    \caption{
      \textit{Greyscale:} SDSS $r$-band image of the A1430 field with galaxies encircled having a spectroscopic redshift close to the cluster redshift.   The main cluster component and the subcomponent are in average separated in redshift.   The colours mark different redshift intervals corresponding to the discussion§ in Sec.\,\ref{sec::redshift_separation} : 
      $0.338 - 0.341$ (green),        
      $0.341 - 0.347$ (blue),        
      $0.347 - 0.348$ (magenta),        
      $0.348 - 0.352$ (red),         
      $0.352 - 0.358$ (orange).
      Pale red contours represent the distribution of X-ray emission as seen by \textit{Chandra} to indicate the cluster position.
    }
    \label{fig::sdss_redshift}
  \end{figure}
  
  % A1430 in SDSS
  A1430 has been observed as part of the Sloan Digital Sky Survey (SDSS), see data release 16 (DR16) \citep[][]{2019ApJS..240...23A}, revealing many galaxies with a redshift close to that of the A1430.  The galaxy at (RA: 11h59m14.9s, DEC: 49d47m48.1s) with a magnitude 17.0 in $i$-band in the most recent data release (DR16) is the BCG, as already classified by \citet{2010ApJS..191..254H}. The second ranked galaxy in the cluster region has a magnitude of 17.7 in $i$-band and is located at (RA: 11h59m30.4s, DEC: 49d47m47.2s), i.e., this galaxy is the brightest galaxy of the subcluster. 

  % spectroscopic redshifts
  The SDSS DR16 lists 27 galaxies with a spectroscopic redshift in the range from 0.3 to 0.4 and with a distance of less than 6.75\,arcmin to the BCG, corresponding to less than 2\,Mpc at the redshift of the cluster.   Fig.\,\ref{fig::sdss_redshift} shows the spatial distribution of the galaxies located close to the X-ray emission and in the redshift range from 0.334 (blue) to 0.360 (red), i.e., close to the cluster redshift. 
  
  % ======= FIGURE: radio galaxies  ======================
  \begin{figure*}[t]
    \centering
    \includegraphics[width=0.32\textwidth]{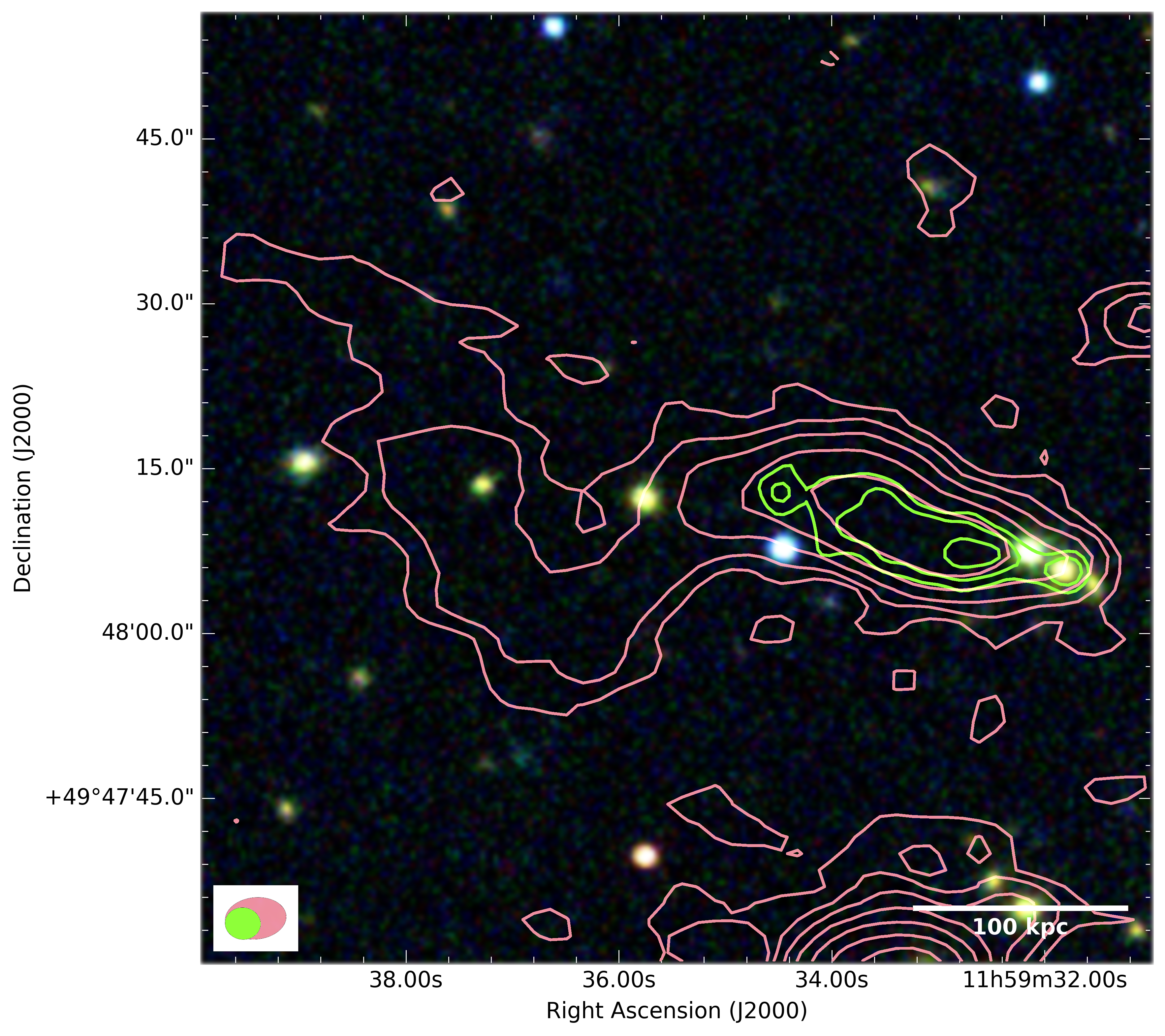}
    \includegraphics[width=0.32\textwidth]{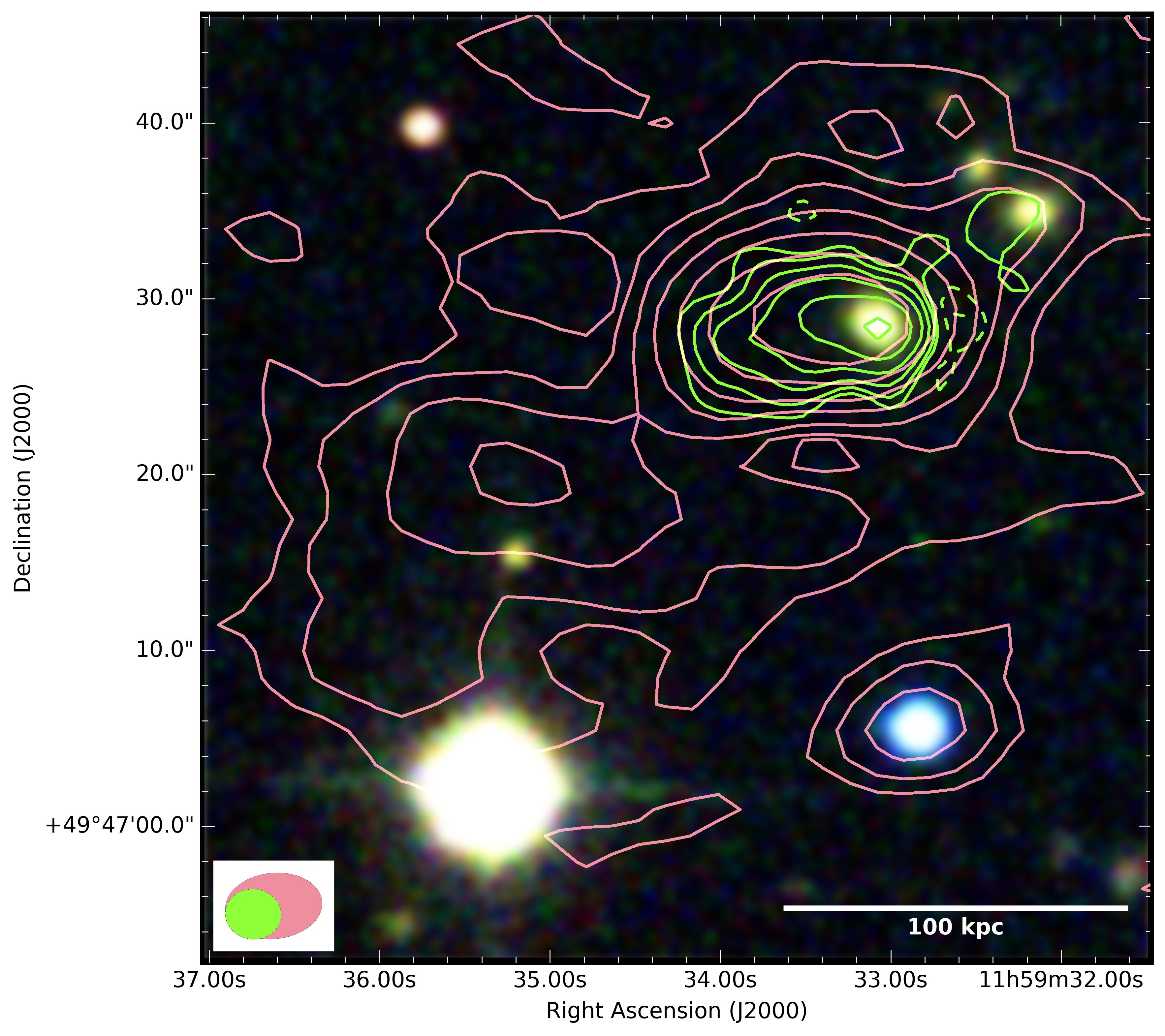}
    \includegraphics[width=0.34\textwidth]{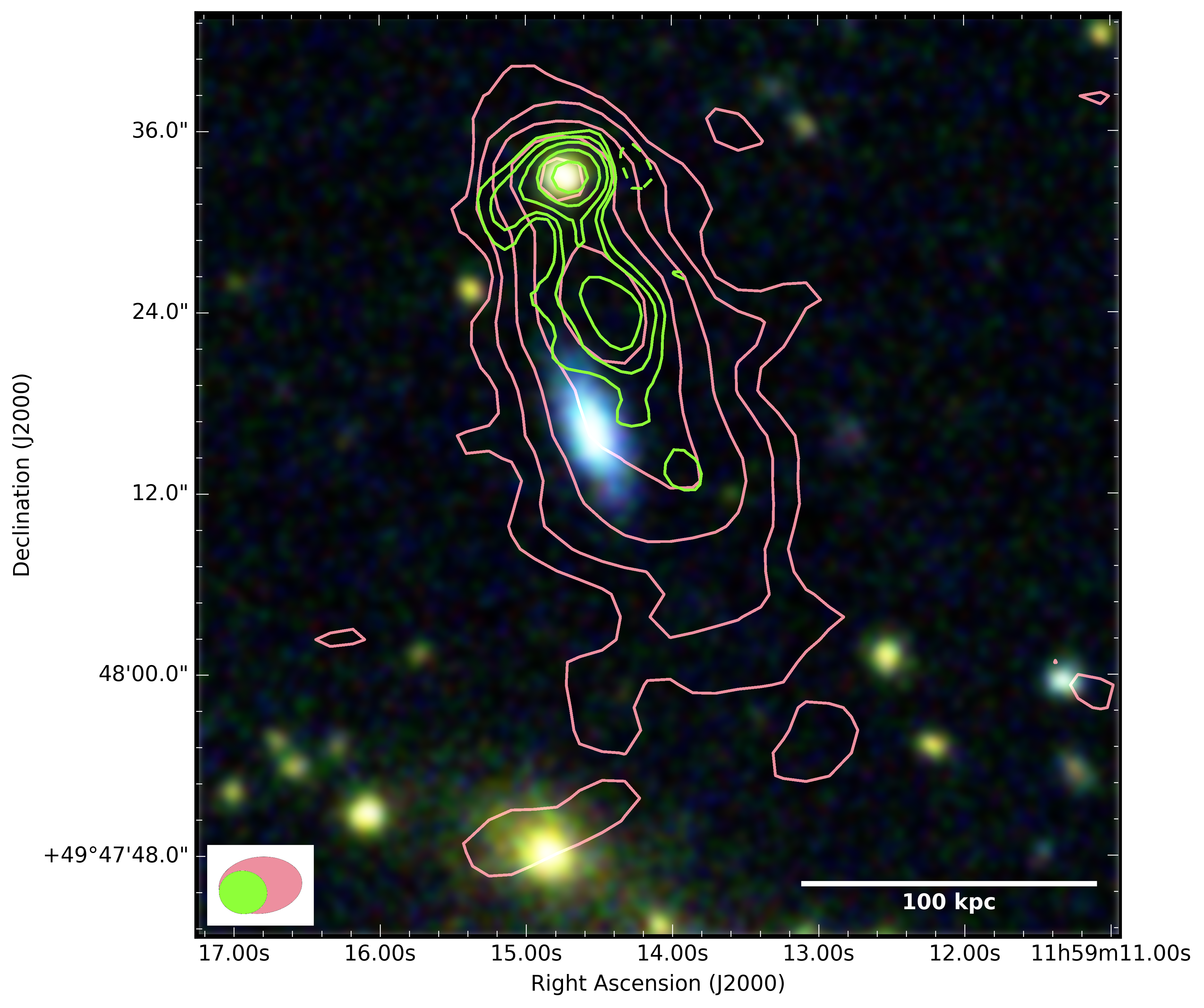}
    \caption{
      Radio galaxies embedded in the diffuse emission in A1430, namely sources a, b, and c (from left to right). \textit{Red contours:} LOFAR HBA high-resolution image achieved using {\tt Briggs} weighting $-1.0$. The noise level is $\sigma_{\text{rms}}=65$\,\textmu Jy\,beam$^{-1}$ with a restoring beam of $5^{\prime\prime}\times4^{\prime\prime}$ and a position angle of $-84^{\circ}$. Contour levels start from $3\sigma_{\text{rms}}$ and are spaced by a factor of $2$. \textit{Green contours:} VLA B-configuration image. The noise level is $\sigma_{\text{rms}}=35$\,\textmu Jy\,beam$^{-1}$ with a restoring beam of $3^{\prime\prime}\times3^{\prime\prime}$ and a position angle of $83^{\circ}$. Contour levels start from $4\sigma_{\text{rms}}$ and are spaced by a factor of $2$. The corresponding beam sizes are depicted in the lower left corner of the images. \textit{Background colorscale:} SDSS DR15 RGB-color composite image using the $g$-,$r$-, and $i$-band, respectively.
      }
    \label{fig::radio_galaxies}
  \end{figure*}

% ==================================================================
% 
%   SECTION:  Galaxy cluster merger and diffuse emission
% 
% ==================================================================

\section{Galaxy cluster merger and diffuse emission}
\label{sec::results_and_discussion}

% ===== SUBSEC :: Two component system =============================
% 
\subsection{A1430: a two-component system}  
\label{sec::two_component_system}

  % two components in X-ray 
  The X-ray surface brightness distribution, as obtained from the \textit{XMM-Netwon} observations, see Fig.\,\ref{fig::XMM}, clearly reveals that A1430 consists of two components, namely the main cluster A and the smaller subcluster B.   The distribution of galaxies listed in SDSS with a spectroscopic redshift corroborates that the cluster has two components, see Fig.\,\ref{fig::sdss_redshift}, even if the number of galaxies is small.   The separation of the two components in the plane of the sky is about $3.1\arcmin$, i.e., about 930\,kpc at the redshift of the cluster.   If both components are at the same distance then they are clearly undergoing a merger.   However, it is also possible that the two components are separated in the radial direction and are not undergoing a merger.  
  
  % X-ray luminosities for A and B
  %  E(z)  = sqrt( OmM (1+z)^3 + OmL )  
  %        =  sqrt(  0.3 * 1.356^3 + 0.7 )  = 1.203 
  %  h_70    = 1 
  As a first step of our analysis, we wish to estimate the X-ray luminosity of each component separately. To this end, we measured the surface brightness in regions which are dominated by one of the components and exclude the interaction --or overlapping-- region.  Moreover, the regions affected by point sources have been excluded. The profile of both components has been approximated with a $\beta$-profile.   This allows us to extrapolate the enclosed luminosity $L_X(<R)$ to regions where the surface brightness is too low to be measured in the \textit{XMM-Newton} observations.   Commonly, the X-ray luminosity is given in the radius $R_{500}$, i.e., the average mass density enclosed is 500 times the critical density in the universe.   As a proxy for modelling the mass distribution in each component, we estimate $R_{500}$ by utilising the luminosity-radius relation $R_{500}(L_X)$ given by \citet{2013A&A...555A..30B}:
  \begin{equation}
      R_{500}
      = 
      \frac{0.957 \, 
      L_{X,500}^{0.207}}{\sqrt{\Omega_m\,(1+z)^3+\Omega_\Lambda}} \, 
      \left( 
        \frac{H_0}{70\,{\rm km\,s^{-1}\,Mpc^{-1}}}
      \right)^{-0.586}  \: 
      ,
  \end{equation}
  where $R_{500}$ is in units of Mpc and $L_{X,500}$ measured in the $0.1-2.4$\,keV band in units of $10^{44}\,\rm erg\,s^{-1}$.
  By demanding that the luminosity of both cluster components A and B should follow the luminosity-radius relation, we can provide $R_{500}$ for each component. We find that components A and B have radii $R_{500,{\rm A}}=3.9\,\arcmin$ and $R_{500,{\rm B}}=3.1\,\arcmin$, respectively. The corresponding X-ray luminosities are $L_{X,500,{\rm A}}=(6.9\pm0.9)\times10^{44}\,{\rm erg \, s^{-1}}$ and $L_{X,500,{\rm B}}=(1.9\pm0.3)\times10^{44}\,{\rm erg \, s^{-1}}$ for the $0.1-2.4$\,keV band. To estimate the uncertainties, we assumed that the radius determination is uncertain by 20\,\%, which is mainly caused by the typical uncertainties in the scaling relations and the uncertainty of the cumulative X-ray surface brightness measurement.
  
  % read luminosity from file :  subcomponent  R=3.1arcsec -->  L_x = 1.9 
  %   20 % uncrtainty in radius   R in [ 2.5, 3.7]   --->  L_xmin = 1.7, L_Xmax= 2.1
  %   uncertainty of xray lum (from files) =  at 3.1 arcsec = 0.16 
  %   total uncertainty Lx for subcomponent   sqrt( 0.2**2 + 0.16**2 ) =  0.3 
  % read luminoisty from file : main component R=3.9arcsec  --> Lx = 6.9
  %   20 % uncertainty in radius  R in [3.12,4.68]  --> L_xmin = 6.15  L_Xmax = 7.45 --> Delta L = (7.45-6.15)/2 = 0.65 %   uncertainty of X-ray lum (from files) =  at 3.9 arcsec = 0.63
  %   total uncertainty Lx for subcomponent   sqrt( 0.65**2 + 0.03**2 ) =  0.9

  % mass estimates
  %  M_500,A = 2.48 x 6.0^0.62  x  1/1.203  = 6.8 x 10^14 M_\odot
  %  M_500,B = 2.48 x 1.9^0.62  x  1/1.203  = 3.1 x 10^14 M_\odot
  % uncertainty estimate : 
  %   use uncertainty in L_x and adopt 20% uncertainty  for scaling relation, includes scaling fact and exp 
  %   This is clearly lower than scatter mentioned in Pratt et al. 2009, however, they mention nongaussian, cool core system
  %   main cluster  0.9/6.9 -->  0.13  ---> uncertainty in M, use exponent 0.62   -
  %      -->  6.8 *  sqrt( 0.2**2 + 0.62*0.13**2)  = 1.5
  %   main cluster  0.3/1.9 -->  0.16  ---> uncertainty in M, use exponent 0.62   -
  %      -->  3.1 *  sqrt( 0.2**2 + 0.62*0.15**2)  =  0.7
  Using a luminosity-mass scaling relation, we can estimate the mass of the two components. From \citet{2013A&A...555A..30B} we find $M_{500,{\rm A}}=(6.8\pm1.5)\times10^{14}\,M_\odot$ and $M_{500,{\rm B}}=(3.1\pm0.7)\times10^{14}\,M_\odot$.  Uncertainty estimates comprise the X-ray luminosity uncertainty and a 20\,\% scatter in the luminosity-mass scaling relation. We note that \citet{2009A&A...498..361P} found a larger scatter for the scaling relations, however, a large deviation was in particular found for cool-core clusters, apparently not the case for A1430. The sum of the two estimated masses is somewhat larger than the total mass derived from the SZ signal, which is $M_{\rm SZ}=(7.6\pm0.4)\times 10^{14} \,\textup{M}_\odot$, however, it is within uncertainty margins.  Moreover, it is not surprising that the two results differ a bit, since for the SZ-based mass estimate of the combined system spherical symmetry is assumed, while we find that the cluster shows a double-peaked morphology. The mass ratio of A1430-A to A1430-B is about 2:1.
  
  % morphological parameters
  As discussed in Sec.~\ref{sec::abell_1430}, the offset between the position of the BCG and the peak of the X-ray surface brightness suggests that the main component A itself is dynamically disturbed. Following \citet{2010ApJ...721L..82C}, we used the morphological indicators within $500\,\rm kpc$, extracted from the point-source subtracted \textit{Chandra} image, to assess the dynamical state of the main cluster A. They found that clusters with and without radio halos are best separated adopting the thresholds for the concentration parameter of $c = 0.2$ and for the centroid shift of $w = 1.2\times 10^{-2}$. For A1430-A, we measure $c_{\rm A1430-A}=0.156 \pm 0.006$ and $w_{\rm A1430-A}=(3.9 \pm 0.9) \times 10^{-2}$ using the X-ray surface brightness distribution obtained with \textit{Chandra}. Therefore, A1430-A belongs, irrespective of the presence of A1430-B, to the class of clusters that typically hosts radio halos. 

  % X-ray temperature 
  To recover the temperature profile of the main cluster component, we extracted spectra in 4 annuli around the centroid of the X-ray emission, ranging from $0.25\arcmin$ to $1.4\arcmin$ (black in Fig.\,\ref{fig::XMM}). At larger radii, we extract spectra in sectors to separate the regions that could be affected by the subcluster (shown in cyan and magenta). We also extract a spectrum in the red annulus to estimate the mean temperature of the subcluster B. We note that the bright emission which may look like the core of the subcluster is clearly associated with a point source in the \textit{Chandra} image and is thus excluded from our analysis. The resulting temperature profile is shown in the right panel of Fig.~\ref{fig::XMM}. The main component A shows a decreasing temperature profile from the centre to the outskirts, with no indication of a cool core, corroborating that the cluster is not relaxed. The mean cluster temperature, measured in a sector excluding the interaction region, amounts to $(7.4\pm0.2)\,\rm keV$. This value is rather high compared to the X-ray temperature expected from the temperature vs. X-ray luminosity correlation, namely $6\,\rm keV$ as computed from the correlation given in \citet{2013A&A...555A..30B}. Moreover, we noticed significant temperature variations at the same radii with the eastern sectors likely affected by the interaction with the subcluster and featuring a larger temperature than the corresponding regions in other directions. Interestingly, the  subcluster B also shows a high temperature of $8.5 \pm 0.7\,\rm keV$.

% ===== SUBSEC :: Radio halo ======================================
%   
\subsection{A radio halo in A1430-A}  
\label{sec::halo}

  % ======= TABLE :: diffuse emission  ============================= 
  \setlength{\tabcolsep}{7pt}
  \begin{table}[!bp]
    \caption{
      Flux densities of the extended diffuse sources in A1430-A (halo) and in A1430-B (Pillow).
      }
    \centering
    \begin{tabular}[]{ccD{X}{\,\pm\,}{-1}}
    \hline 
    \hline
    Object    & Property             & \multicolumn{1}{c}{Value}             \\
    \hline 
     % main cluster & $L_{X,500,{\rm A}}$ & $6.0\times10^{44}\,\rm erg \, s^{-1}$ \\
     %  & $M_{500,{\rm A}}$  &  $ 6.3\times10^{14}\,M_\odot $ \\
       % main cluster & $L_{X,500,{\rm A}}$ & $6.0\times10^{44}\,\rm erg \, s^{-1}$ \\
    % \hline
     %  subcluster & $L_{X,500,{\rm B}}$  & $1.9\times10^{44}\,{\rm erg \, s^{-1}}$ \\
     %  & $M_{500,{\rm B}}$  &  $ 1.9\times10^{14}\,M_\odot $
    \hline
    halo      & $S_{A,144}$          & 21\phantom{.0} X\phantom{1}6\phantom{.4}  \rm\,mJy           \\
    (A1430-A) & $S_{A,1.5}$          & 1.1X\phantom{1}0.4\rm\,mJy            \\
              & $\alpha_{144}^{1.5}$ & -1{.}3X\phantom{1}0.3          \\
    \hline 
    Pillow    & $S_{A,144}$          & 36\phantom{.0}X18\phantom{.4}\rm\,mJy \\
    (A1430-B) & $S_{A,1.5}$          &  1.5X\phantom{1}0.8\rm\,mJy           \\
              & $\alpha_{144}^{1.5}$ & -1{.}4X\phantom{1}0.5          \\
    \hline 
    \hline
    \end{tabular}
    \label{tab::diffusesources}
  \end{table}
  
  % ======= FIGURE :: diffuse emission + X-ray overlay  ================
  \begin{figure*}[htbp]
    \centering
    \includegraphics[width=0.8\textwidth]{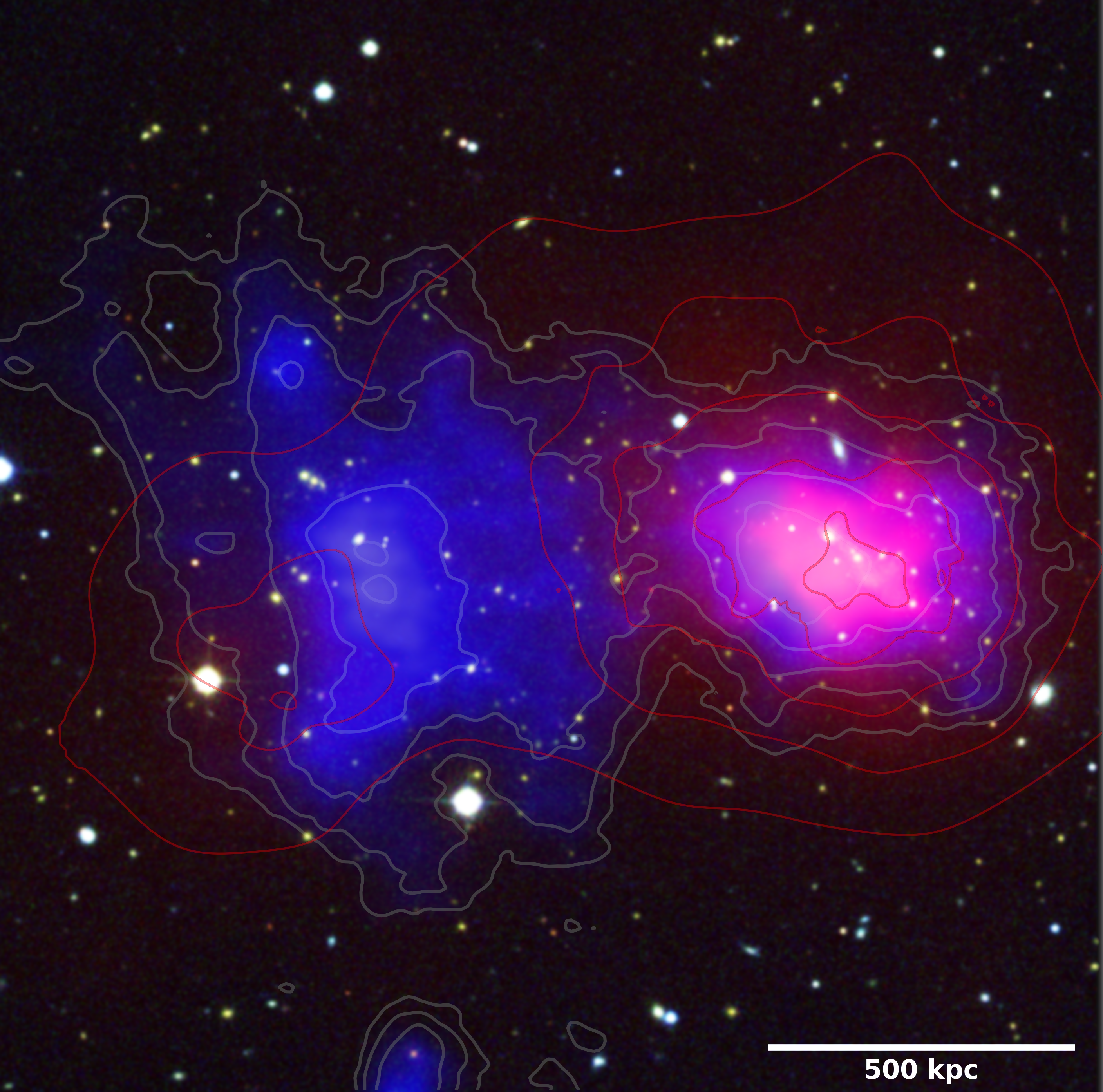}
    \caption{
      LOFAR low resolution, X-ray and optical overlay of the galaxy cluster A1430. The image clearly reveals the halo and large scale diffuse radio emission in the low density ICM, dubbed as Pillow. The image is obtained using Briggs weighting $+0.4$ and \texttt{multiscale clean} after subtraction of sources listed in Tab.\,\ref{tab::compactsources}. The noise level is $\sigma_{\text{rms}}=80$\,\textmu Jy\,beam$^{-1}$ and the restoring beam is $25^{\prime\prime}\times22^{\prime\prime}$ with a position angle of $6^{\circ}$. Contour levels are drawn from 3\,$\sigma_{\text{rms}}$ and are spaced by a factor of $\sqrt{2}$. \textit{Blue colorscale:} LOFAR HBA low-resolution image, see Fig.\,\ref{fig::A1430_subtracted} for image properties. \textit{Red colorscale:} Smoothed \textit{Chandra} X-ray emission in the 0.5\,-\,7.0 keV energy band. \textit{Background colorscale:} SDSS DR15 RGB-color composite image using the $g$-,$r$-, and $i$-band, respectively.}
    \label{fig::diffuse}
  \end{figure*}

  % the LOFAR result
  %   LoTSS iamge discussed on fig LoTSS highres
  % evidence for diffuse emission  
  In the field of A1430, no diffuse emission has been previously reported. For the first time, the LoTSS observations showed evidence for extended, diffuse radio emission in this cluster. As discussed in Sec.\,\ref{sec::uncoverdiffuse},  the bright, tailed radio galaxies and other compact sources in the field have to be subtracted before the diffuse emission can be imaged at a low resolution allowing to determine the flux density and morphology of the diffuse emission. 
  
  % Fig.\,\ref{fig::radio_galaxies} shows high-resolution LOFAR images of the three tailed radio galaxies.  These images have been used to model the radio galaxies and to subtract them from the visibilities before imaging the diffuse emission. The analysis of the halo emission is affected by source c. It is apparently a head-tailed radio galaxy with a rather straight tail.  The source can be reasonably well modelled and subtracted. 
  
  % morphology similar to X-ray, classify as halo
  % LLS : 2.3 arcmin * 60 sec/min * 5 kpc / arcsec
  After subtracting the radio galaxies we re-imaged with a resolution of about 25\,arcsec, see Fig.\,\ref{fig::diffuse}. The main cluster component A clearly shows diffuse emission with a largest linear size of about 700\,kpc and a morphology very similar to the X-ray morphology.  Based on these morphological properties, we classify the diffuse radio emission in A1430-A as a radio halo. 
  
\begin{figure}[tbp]
    \centering
    \includegraphics[width=0.48\textwidth]{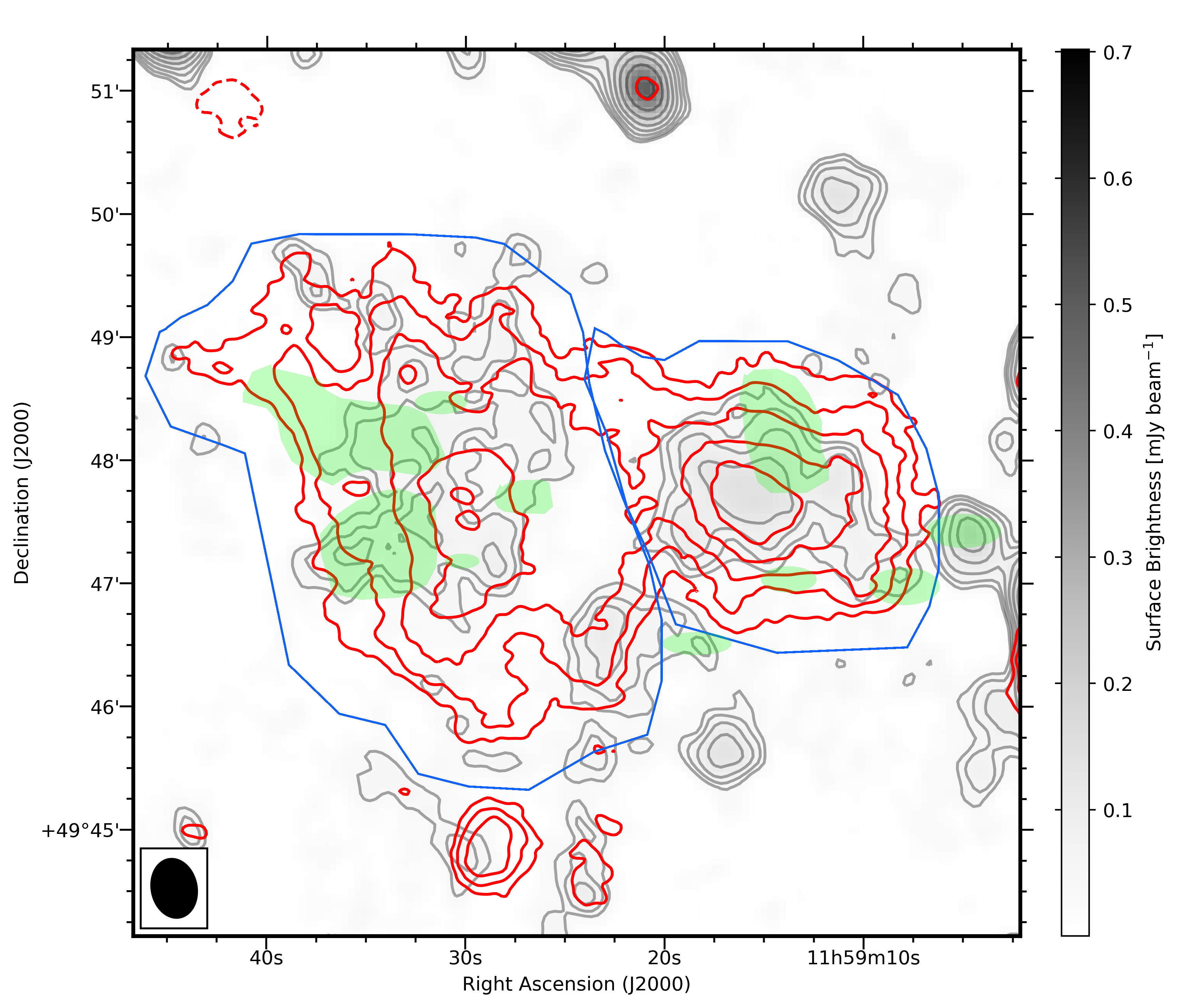}
    \caption{
      Total power low-resolution radio map of A1430. {\it Greyscale:} VLA image using Briggs weighting $+0{.}6$ with sources subtracted. The noise level is $\sigma_{\text{rms}}=13$\,\textmu Jy\,beam$^{-1}$ with a restoring beam of $29^{\prime\prime}\times22^{\prime\prime}$ and a position angle of $14^{\circ}$. {\it Red contours:}  LOFAR HBA image and contours as in Fig.\,\ref{fig::diffuse}. Negative contour lines are drawn with dashed lines. {\it Blue lines:} Regions to indicate which diffuse emission is considered to be part of each component of the cluster. {\it Green areas} Regions where compact and extended sources have been subtracted in the $uv$-plane for both the VLA and the LOFAR image.}
    \label{fig::A1430_subtracted}
  \end{figure}

  % radio power 
  %  z = 0.356 
  %  estimate spectral index  alpha = -1.1    
  %  P_nu = 4 pi D_L^2 (1+z)^(-(1+alpha)) S_nu
  %  D_L(z=0.356)  =  1894 Mpc   (H0=70 OmMatter=0.3)
  %  S_nu = 0.016 Jy 
  %  P_144  = 4 pi (1894 Mpc * 3.086x10^24 cm/Mpc)^2
  %             x (1.356)^0.2  x 0.021 x 10^(-23) erg/s/Hz/cm2 
  %         = 4 pi x 5.845^2 x 1.356^0.2 x 0.021  x  10^31 erg/s/Hz
  %         = 9.6 x 10^31 erg/s/Hz
  %         = 9.6 x 10^24 W / Hz
  %  P_1.5  = 4 pi (1894 Mpc * 3.086x10^24 cm/Mpc)^2
  %             x (1.356)^0.2  x 0.0011 x 10^(-23) erg/s/Hz/cm2 
  %         = 4 pi 5.845^2 1.356^0.1 0.0011  x  10^31 erg/s/Hz
  %         = 0.5 x 10^31 erg/s/Hz
  %         = 0.5 x 10^24 W / Hz
  For the halo, we measure a flux density of $S_{\rm halo, 144} = 21 \pm 5 \,\rm mJy$; see Fig.\,\ref{fig::A1430_subtracted} for the area of the measurement. The flux density corresponds to a rest-frame luminosity $P_{\rm halo, 144} = 1.0 \pm 0.3 \times 10^{25} \,\rm W \, Hz^{-1} $.

  \citet{2020arXiv201102387V} reported a higher flux density for the halo, namely $29.8 \pm 6.6 \,\rm mJy$. There are two reasons for this difference: \citet{2020arXiv201102387V} are using the integrated flux densities obtained from fitting exponential profiles. This increases the flux density since the halo is extrapolated to the very low surface brightness regime. Moreover, with the method applied here we subtract for the radio galaxy c 6\,mJy more than with the $uv$-cut method, see Sec.\,\ref{sec::uncoverdiffuse} for a discussion.

  % ======= FIGURE :: LOFAR diffuse + VLA + regions  ==========
  
  % VLA results  
  % spectral index
  % S_1 / S_2 = ( nu_1 / nu_2 )^alpha
  % alpha = log( S_1 / S_2 ) / log ( nu_1 / nu_2 )
  % nu_1 = 1500 MHz, nu_2 = 144 MHz,    S_1 = 1.1 mJy  S_2 = 16.5 mJy
  %  alpha = -1.2 
  % Delta alpha 
  %      = sqrt( sum_i ( Delta S_i / S_i )^2 ) / | ln( nu_1 / nu_2 ) |
  %      = sqrt( (5.2/16.5)^2 + (0.4/1.1)^2 ) / 2.34 
  %      = 0.2 
  There is also evidence for diffuse emission in the galaxy cluster in the VLA L-band image for both the main cluster component A and the subcomponent B, see Fig.\ref{fig::VLAimagea}. Similar to the LOFAR image, the bright radio galaxies, see Fig.\,\ref{fig::radio_galaxies}, and fainter compact sources need to be properly subtracted at 1.5\,GHz before imaging the low surface brightness diffuse emission at a low resolution.  The medium-resolution compact source subtracted image is shown in Fig.\,\ref{fig::VLAimage}. The image is produced at $29\arcsec\times 22\arcsec$ resolution using C- and D-configurations data and using {\tt robust=0.6}. The halo is clearly detected and its morphology basically agrees with the one recovered with the LOFAR at 144\,MHz, see Fig.\,\ref{fig::A1430_subtracted}. The halo is less extended than in LOFAR, likely due to a lower signal-to-noise ratio in the VLA image. At 1.5\, GHz, we measure a flux density of $S_{\rm halo,1.5} = 1.1 \pm 0.4 \,\rm mJy$ for the halo in A1430-A.  The flux density uncertainty is estimated according to Eq.\,\ref{eq::fluxerror}.  The low signal-to-noise ratio, which determines to which extent the surface brightness distribution can be recovered, and the remaining compacted sources, which are too faint to be subtracted, might affect the halo flux density measurement beyond the given uncertainty.  
 
  % VLA radio power  and spectral index 
  The corresponding rest-frame radio power is $P_{\rm halo, 1.5} = 5 \pm 2 \times 10^{23} \,\rm W \, Hz^{-1}$.  The flux densities, determined from the VLA and LOFAR low resolution images, correspond to a spectral index $\alpha_{144}^{1.5} = -1.3 \pm 0.3$, a value which is typical for radio halos \citep{feretti_2012}.

  With the LOFAR and VLA results at hand, we reconsider the scenario that the diffuse emission found in the main cluster  might actually originate from an extended lobe of the radio galaxy c. As found above, the halo morphology is at both frequencies, LOFAR and VLA, very similar to the X-ray morphology and the extended emission shows a rather uniform spectral index of $\alpha_{144}^{1.5} = -1.3 \pm 0.3$. This disfavours that this emission is an old lobe of the radio galaxy, instead it corroborates the classification as radio halo. 
  
\begin{figure}[htbp]
    \centering
    \includegraphics[trim = 23 2 19 0, clip, scale=0.78]{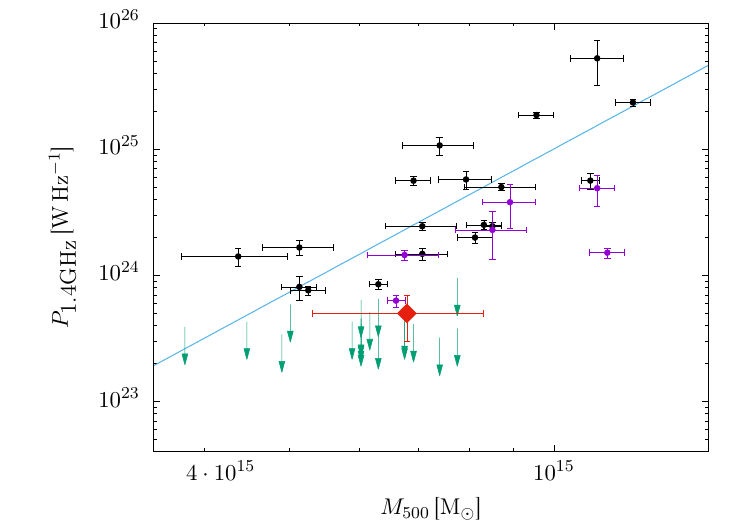}
    \caption{
      Radio power at 1.4\,GHz vs. cluster-mass relation. \textit{Black filled dots:} Radio halo galaxy clusters. \textit{Violet filled dots:} Galaxy clusters possessing ultra-steep spectrum radio halos. \textit{Green arrows:} Upper limits of potential radio halos (non-detections). The red diamond indicates the radio halo in Abell\;1430-A. The best-fit relation for radio halos is drawn as blue line. Plot and data are adapted from \citet{2013ApJ...777..141C}.
      }
    \label{fig::radio-xray-correlation}
  \end{figure}

  % typical halo radio powers of clusters similar to A1430
  %      cassano+ 2013 sample
  %     refer to Fig. correlation
  %    A1430 is underluminous, as A1451 and Z0634, as classified by Cuciti+ 2018
  Halos in galaxy clusters with an X-ray luminosity similar to the one in A1430 show a significant variety of rest-frame radio luminosities, see, e.g., the clusters compiled by \citet{2013ApJ...777..141C}. For instance, the halo in Abell\,1995 has a radio power at 1.4\,GHz of $ 1.7 \pm 0.2 \times 10^{24} \,\rm W \, Hz^{-1}$ \citep{2009A&A...507.1257G}, the one in Abell\,545 has $1.4 \pm 0.2 \times 10^{24} \,\rm W \, Hz^{-1}$ \citep{2003A&A...400..465B}, the one in Abell\,773 has $1.48 \pm 0.16 \times 10^{24} \,\rm W \, Hz^{-1}$ \citep{2001A&A...376..803G} and the one in Abell\,2256 has $8.1 \pm 1.7 \times 10^{23} \,\rm W \, Hz^{-1}$ \citep{2006AJ....131.2900C}. Compared to the rest-frame radio power of these halos, the one in A1430-A is less luminous, see Fig.\,\ref{fig::radio-xray-correlation}. Recently, \citet{2018A&A...609A..61C} reported in the clusters Abell\,1451 and ZwCl\,0634+47, which show an X-ray luminosity similar to the one of A1430-A, two radio halos with a radio power of $ 6.4 \pm 0.7 \times 10^{23} \,\rm W \, Hz^{-1}$ and $ 3.1 \pm 0.2 \times 10^{23} \,\rm W \, Hz^{-1}$, respectively.  These halos are underluminous with respect to the luminosities of radio halos in samples compiled before.  According to the nomenclature introduced by  \citet{2011ApJ...740L..28B} and used by \citet{2018A&A...609A..61C}, the halo in A1430-A as observed at 1.5\,GHz belongs to the class of underluminous halos. 
  
  % speculate origin of underluminous:
  %    selection bias (faint halos difficult to detect)
  %    minor merger, not likely since radio and Xray SB so similar
  %    off-state, hadronic,  not likely, since there are already fianter upper limits 
  %  most likely: A1430 is underluminous since in transition phase, end state of halo
  There is no clear understanding yet why halos, such as the one in A1430-A, are underluminous. There could be a selection bias, since recovering halos with a total flux density of the order of 1\,mJy is still challenging, hence, many faint halos are possibly still undetected. As a consequence, samples compiled so far may tend to comprise only brighter halos at least for less massive galaxy clusters.  Irrespectively of such a selection bias, it becomes evident that halos in clusters with similar X-ray luminosity show a variety of radio powers.  We can only speculate about the reason for this variety.  Clearly, there must be some transition phase in the evolution of a cluster from showing no halo, presumably when the cluster is relaxed, to showing a `fully powered' halo, presumably as a result of a major merger.  Therefore, we expect to observe some clusters with halos showing an intermediate radio power. Moreover, \citet{2018A&A...609A..61C} speculate that underluminous halos are caused by minor mergers, where only part of the ICM becomes turbulent, hence the halo would be restricted to a subvolume of the cluster. For A1430-A the radio and X-ray surface brightnesses correlate nicely, therefore we consider the scenario of a minor merger as unlikely.  Finally, \citet{2018A&A...609A..61C} speculate that underluminous halos might reflect the off-state of halos, where the hadronic halo becomes visible.  Since there are already upper limits at lower flux densities, it is not likely that the halo in A1430-A is a hadronic one. Since we consider neither the off-state or the minor merger scenario likely for A1430, we speculate that A1430 is actually in the transition phase between the off-state and a `fully powered' halo. The absence of a cool core, see Fig.\,\ref{fig::XMM}, may indicate that we actually witness the late state of a merger, in which the ICM has been mixed and had no time to form a cool core yet.

  % ======= FIGURE :: radio power vs x-ray lum correlation  ==========

% ===== SUBSEC :: radio Pillow in A1430-B ==================
%  
\subsection{Diffuse emission related to subcluster A1430-B}
\label{sec::diffuse_A1430-B }
  
  % ======= FIGURE :: VLA diffuse emission image  =========
  \begin{figure}[htbp]
    \centering
    \includegraphics[width=0.5\textwidth]{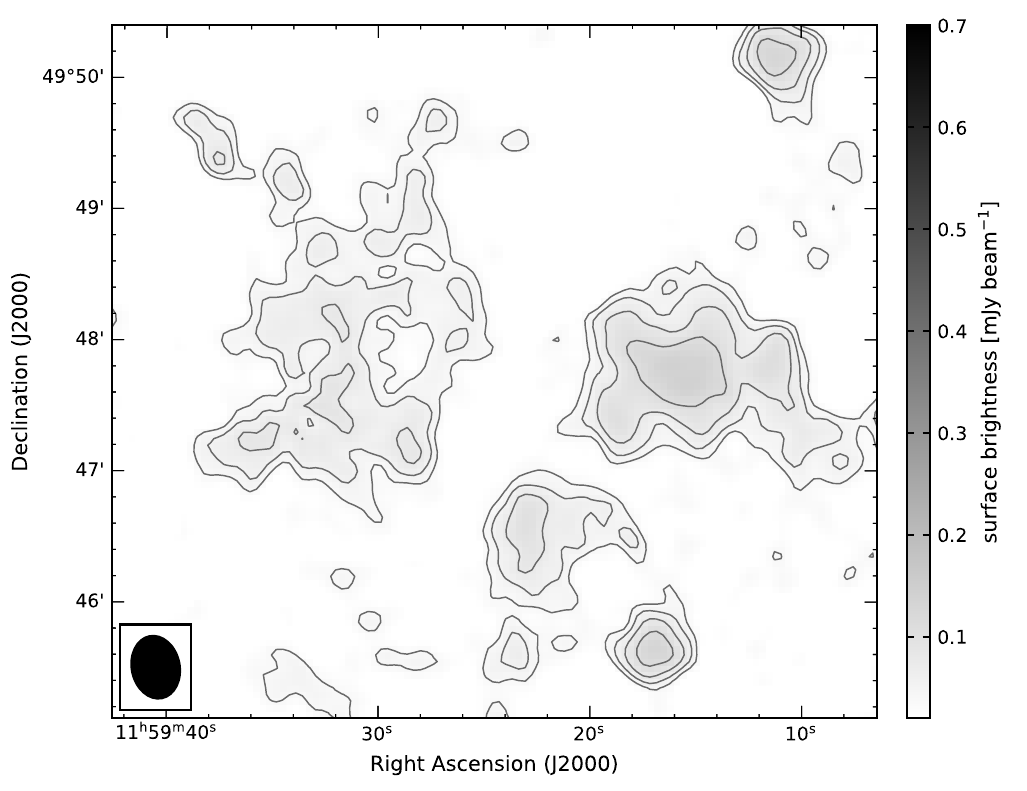}
    \vspace{-0.6cm}
    \caption{ 
      VLA 1-2\,GHz point-source subtracted image of the cluster A1430. We subtracted all point sources labelled in Fig\,\ref{fig::LoTSS-highres} within the green areas shown in Fig.\,\ref{fig::A1430_subtracted}. The restoring beam size of the image is $29\arcsec\times22\arcsec$ and the noise level is $\sigma_{\text{rms}}=13$\,\textmu Jy\,beam$^{-1}$. Contour levels are drawn at $\sqrt{[1,2,4,8,\dots]}\,\times\,3\,\sigma_{{\rm{ rms}}}$. No negative contours are present.
      }
    \label{fig::VLAimage}
  \end{figure}

  % a large emission region related to A1430-B
  The most striking radio emission feature in A1430 is large diffuse emission apparently related to the subcluster B, see Fig.\,\ref{fig::diffuse}. Due to its morphology, we dub it in the following `Pillow'. It is located to the west of the two radio galaxies A and B.

  Because the tails of the two radio galaxies are directed towards the east, the majority of the Pillow emission does not overlap with these radio galaxies, only part of the emission of the Pillow to the east is difficult to recover due to the radio galaxy subtraction.
  The emission region has a large extent, at least about 1.2\,Mpc in diameter. The surface brightness of the emission appears to be quite smoothly distributed, without any evident outer edges, large filaments, or other discontinuities.
  
  % flux density and radio power
  %   substitute flux values in halo section
  The flux density of the emission amounts to $S_{\rm Pillow,144} = 36 \pm 18 \, \rm mJy$, see Fig.~\ref{fig::A1430_subtracted} for the region where the flux density has been measured. The flux density corresponds to a rest frame luminosity of $P_{\rm A, 144\,\rm MHz} = 1.7 \pm 0.8 \times 10^{25} \,\rm W \, Hz^{-1} $.
  
   One may speculate that the Pillow actually originates from old lobes of one or more radio galaxies. Candidates would be evidently the galaxies a and b. However, we consider this very unlikely for the following reasons: Firstly, galaxy a has a clear head-tail morphology with the tail to the east, suggesting that the galaxy is falling into the subcluster and making it unlikely that luminous lobe emission is found to the east. Secondly, also galaxy b shows a head-tail morphology with a tail to the south-east, indicating that the galaxy is falling into the cluster. Moreover, the Pillow is rather luminous and its surface brightness roughly agrees with the X-ray surface brightness, except towards the east, where the presence of the radio galaxies a and b prevent us from determining its morphology. In addition to the orientation of a and b, also the morphology of the Pillow and its correlation to the X-ray surface brightness disfavour that the Pillow originates from old lobes.

  % L-band VLA results 
  % spectral index
  % S_1 / S_2 = ( nu_1 / nu_2 )^alpha
  % alpha = log( S_1 / S_2 ) / log ( nu_1 / nu_2 )
  % nu_1 = 1500 MHz, nu_2 = 144 MHz,    S_1 = 1.5 +/- 0.8  mJy  S_2 = 30 pm 17 mJy
  %  alpha = -1.278 
  % Delta alpha 
  %      = sqrt( sum_i ( Delta S_i / S_i )^2 ) / | ln( nu_1 / nu_2 ) |
  %      = sqrt( (17/30)^2 + (0.8/1.5)^2 ) / 2.34 
  %      = 0.33  
  In the L-band low resolution image the Pillow is clearly visible as well, see Fig.\ref{fig::VLAimage}. We measure a flux density of $S_{\rm Pillow,1.5} = 1.5 \pm 0.8\,\rm mJy$ after source subtraction. This implies that the Pillow has a moderately steep spectrum with a spectral index $ \alpha_{144}^{1.5} = -1.4 \pm 0.5 $. We note that the subtraction of the compact sources make up a large share of the flux density uncertainty. Moreover, the extended emission with a surface brightness only a few times above the noise level is notoriously difficult to deconvolve in interferometric data. To definitely determine the surface brightness distribution of the Pillow substantially deeper observations are required. However, the detection of patches of diffuse emission, especially at the location where the peak of the Pillow has been found in the LOFAR image, clearly indicates a spectral index flatter than about $-2$, disfavouring that the emission is caused by fossil plasma ejected by an active galactic nucleus (AGN) some time ago. The spectral index rather suggests that the Pillow is caused by turbulence or shocks in the ICM as it is the case for halos or relics, respectively.

  % a possible relic ? 
  %   not likely since: no outer edge, no spectral steepening, nor polarized emission
  %   however, possibility cannot be fully ruled out
  The emission here is located at the cluster periphery of A1430-A. A conceivable scenario for the origin of the Pillow is that A1430-A underwent a major merger some time ago, the same merger which causes the halo in A1430-A, and the Pillow is actually a radio relic caused by a large merger shock. The location of the Pillow would be roughly in agreement with the elongation of A1430-A, indicating the merger axis.  To not prevent a merger shock front from forming, A1430-B would have to have some offset along the line of sight w.r.t. to A1430-A. However, the Pillow does not show the outer edge typical for relics, hence, the Pillow morphology differs significantly from the one common for relics. The LOFAR and the VLA images revealed a similar morphology of the Pillow, excluding a clear spectral steepening, which would be characteristic for a relic. We note that the diffuse emission in Abell\,2256 \citep{2012A&A...543A..43V,2014ApJ...794...24O} also shows a morphology very different from `typical' relics and is considered as the rare situation where a relic is believed to be seen face-on. The relic in Abell\,2256 is highly polarised at 1.5\,GHz, reaching values as high as 70\,\% \citep{2014ApJ...794...24O}. We do not find evidence for polarised emission for the Pillow in the VLA data, however, due to the low surface brightness of the Pillow we can only rule out a fractional polarisation as high as a few ten percent. To summarise, we do not find any confirming evidence for a classification as relic, in particular neither an outer edge coinciding with a jump in X-ray surface brightness nor an apparent gradient in the spectral index, nor polarised emission. We therefore consider it as unlikely that the Pillow is a radio relic related to a merger shock front, however, we cannot fully rule out that possibility.

  % turbulence in a connecting filament 
  %   similar to A399 and A401 pair, to 1758N and A1758S pair
  %   Pillow may actually be a radio bridge seen in projection
  %   refer to Brunetti and Vazza 2020 second order Fermi in filament
  If the Pillow is not a relic, it most likely originates from turbulence, similar to radio halos.  Recently, radio emission has been reported for the filaments which connect the two clusters pairs Abell\,399 -- Abell\,401  \citep{2019Sci...364..981G} and Abell\,1758N -- Abell\,1758S \citep{2018MNRAS.478..885B,2020MNRAS.499L..11B}. This may indicate that in an early stage of a galaxy cluster merger, when the two clusters are still approaching each other, the medium in the connecting filament gets sufficiently disturbed to generate low Mach number shocks and to accelerate thermal or mildly relativistic electrons.  \citet{2020PhRvL.124e1101B} showed that those radio bridges, extending on scales larger than clusters, may originate from second order Fermi (re-)acceleration of electrons interacting with turbulence originating from the complex dynamics of substructures in the filaments. The spectral index we found for the Pillow indeed agrees with the required steepness of $-1.3$ to $-1.5$ or steeper for the proposed acceleration mechanism. A plausible scenario for the origin of the Pillow therefore is, that A1430-A and A1430-B are significantly separated along the line of sight and we actually observe the radio emission of a connecting filament in projection.  To test this scenario,  deeper observations would be crucial to firmly determine the spectral properties of the Pillow.

  % X-ray: 
  %   rather high temperatures in Pillow region
  %   high T in cyan region: speculate merger shock from merger in A 
  Interestingly, the X-ray temperature in the sector between the two clusters, see Fig.\,\ref{fig::XMM}, left panel (magenta region), is much higher than at the same distance from the cluster centre in other directions and the same distance from the main cluster, see Fig.~\ref{fig::XMM}, right panel. It suggests that adiabatic compression, shock dissipation or turbulence resulting from the interaction of the subcluster with the main cluster may have already increased the temperature in that region, possibly similar to the situation in the cluster pair Abell\,1758N and Abell\,1758S \citep{2018MNRAS.478..885B}.  Interestingly, also the region to the north-east of A1430-A (cyan region) shows a rather high temperature. We speculate that this is related to a merger shock front originating from the merger which caused the halo in A1430-A and is not related to the Pillow emission.

  % ======= FIGURE :: The300 simulated cluster X-ray SB and temp  =====
  \begin{figure*}[tbp]
    \centering 
    \hspace{0.3cm}
    \includegraphics[width=0.48\textwidth]{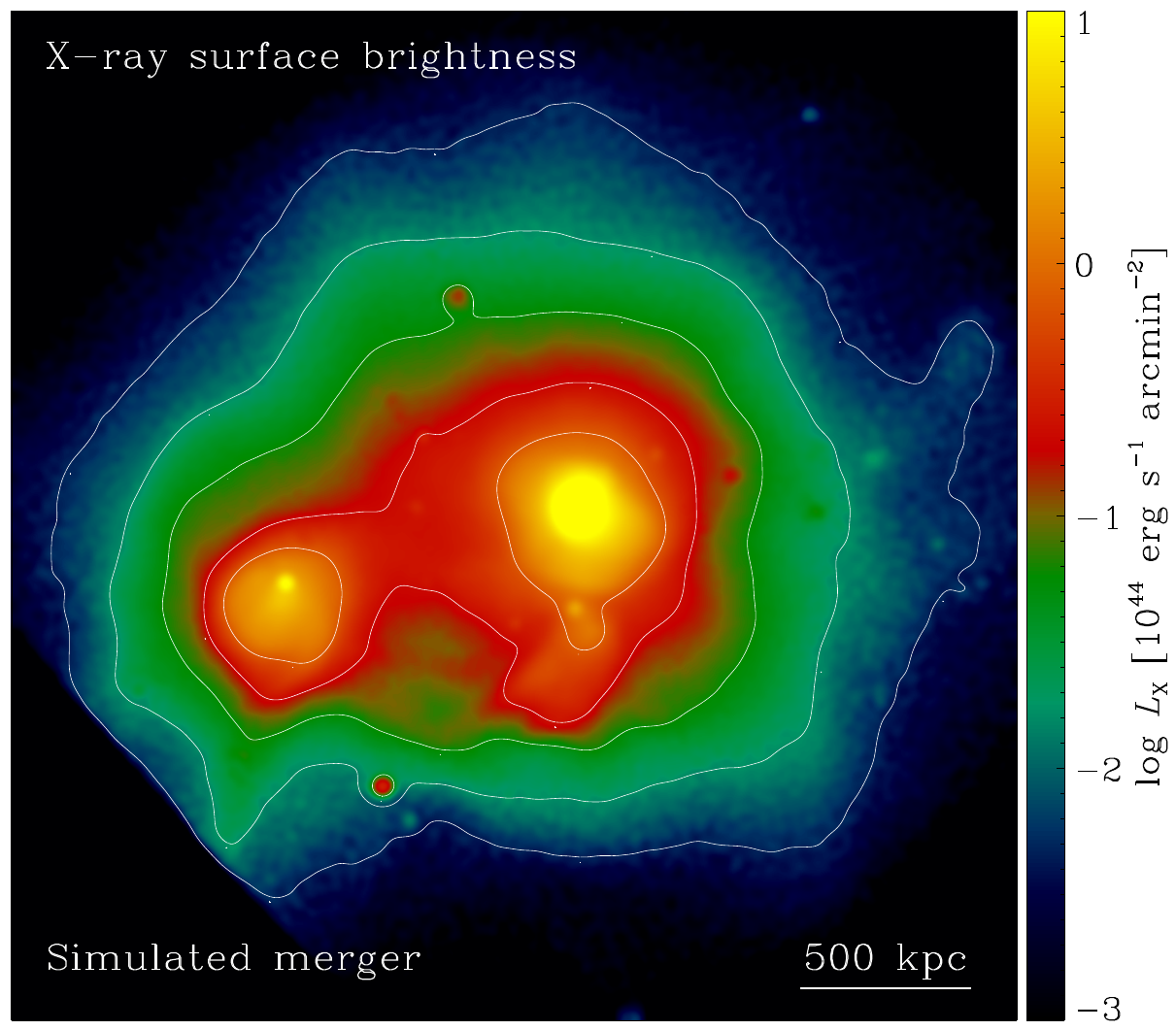}
    \hspace{0.15cm}
    \includegraphics[width=0.48\textwidth]{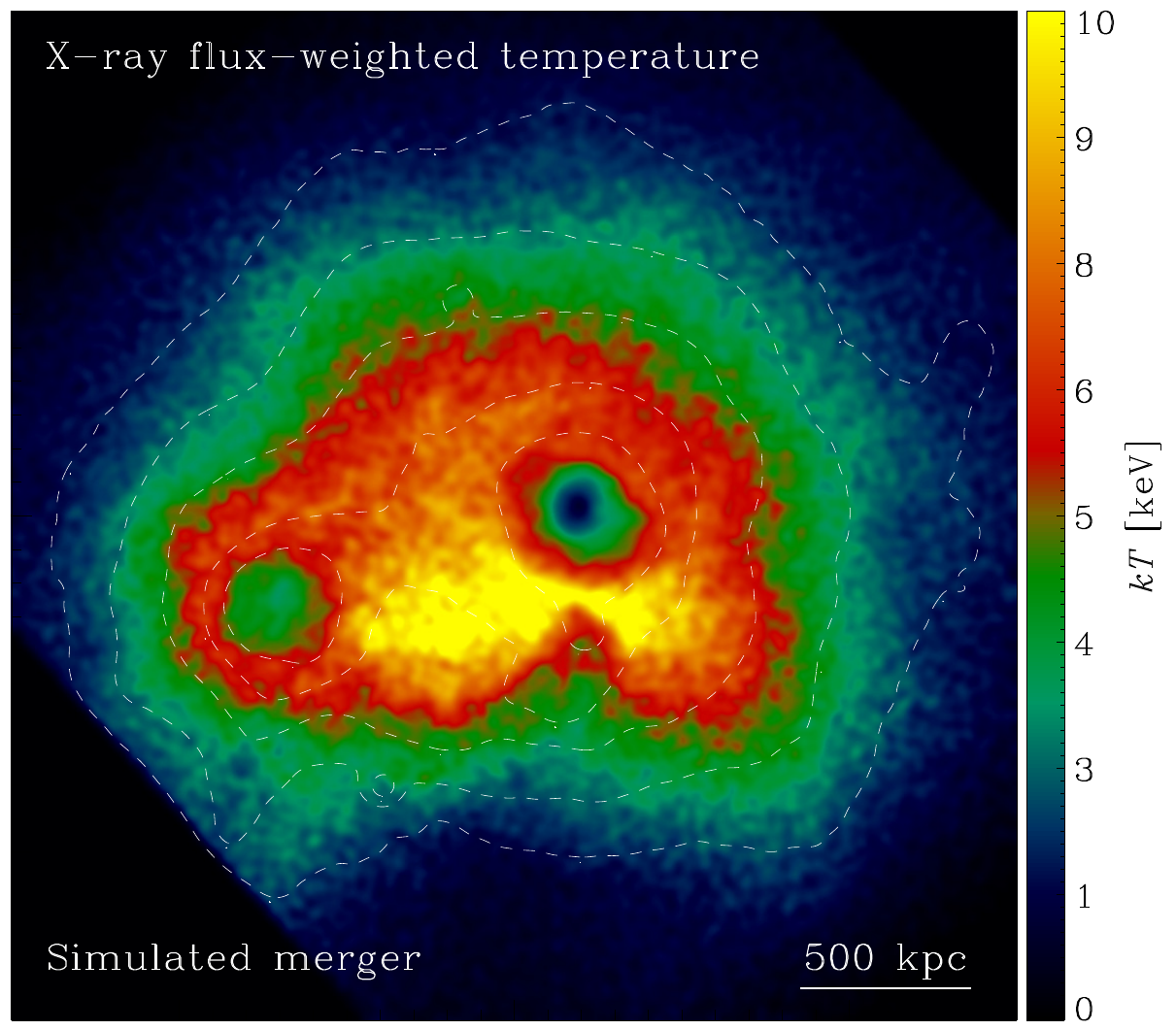}
    \caption{
      Example of a simulated cluster merger candidate chosen to match observed A1430's properties smoothed with a Gaussian of $3\arcsec$ width. 
      {\it Left panel:} X-ray surface brightness distribution. Merging clusters are clearly seen as two distinct peaks. Solid contours indicate logarithmic X-ray surface brightness levels of $42,42.5,43,43.5$ and $44$ in units of ${\rm erg}\,{\rm s}^{-1}\,{\rm arcmin}^{-2}$. 
      {\it Right panel:} X-ray flux-weighted temperature. For comparison, surface brightness contours are also shown as dashed lines. Gas temperature increases between the clusters as a result of the interaction.  
    }
    \label{fig::sim_merger_1}
  \end{figure*}

% ===== SUBSEC :: redshift separation A and B  ==================
% 
\subsection{Redshift separation of A1430-A and B}  
\label{sec::redshift_separation}

  % sdss redshift distribution
  %   hint for redshift difference between A and B
  Fig.\,\ref{fig::sdss_redshift} shows the SDSS $r$-band image where 19 out of 27 galaxies are marked with a colour-coding that represents redshift subintervals. Though there is a concentration of the red symbols towards the right-hand side and of the blue ones towards the left-hand side, there is no clear spatial separation of the redshifts.   Therefore, substructures should be additionally identified in redshift space. We decided for an approach that is based on a combination of the redshift distribution and the positions of the BCGs.

  % three candidate BCGs
  The `Optical Catalog of Galaxy Clusters Obtained from an Adaptive Matched Filter Finder Applied to SDSS DR6' \citep{Szabo_2011} lists three BCG candidates in the cluster.  For the candidate BCGs c1, c2 and c3 redshifts of $0.3499$, $0.3508$ and no spectroscopic redshift, respectively, were given. In SDSS DR16, a spectroscopic redshift of $0.34436$ is reported for BCG\,c3. In Fig.~\ref{fig::Chandra} the candidate BCGs are marked with orange circles.

  % first model 
  % BCG c1 and c3 represent center of potential wells
  %  vel separation / rad distance from avarage redshift in A and B 
  %  sigma --> Delta z  :  v =c Deltaz --> Delta z = v/c = 0.00333
  %   z_BCGc1 = 0.3499   --> interval :  0.34657 - 0.35223
  %   z_BCGc3 = 0.34436  --> interval :  0.34104 - 0.34769
  % center redshift between both components: 0.34713
  %   limits 4 * v/c around center redshift :  z = [0.334, 0.36]
  We assume that the brightest galaxy in the sample, BCG\,c1, represents the bottom of the gravitational potential and its redshift is representative of the cluster. Moreover, we assume a typical velocity dispersion of $\sigma_z = 1000$ km s$^{-1}$ and classify all sample galaxies within the redshift range from $ z_{\rm BCG\,c1} - \sigma_z $ to $ z_{\rm BCG\,c1} + \sigma_z$ as members of cluster A1430-A (magenta and red symbols in Fig.\,\ref{fig::sdss_redshift}). According to its redshift, the BCG\,c2 would belong to cluster A1430-A, but the BCG\,c3 would not. We consider BCG\,c3 as the brightest galaxy in A1430-B with candidate members that are selected with redshift from $ z_{\rm BCG\,c3} - \sigma_z $ to $ z_{\rm BCG\,c3} + \sigma_z $ (blue and magenta symbols). The redshift distributions associated with A1430-A and B overlap marginally (magenta symbols).   Finally, galaxies with redshift smaller $ z_{\rm BCG\,3} - \sigma_z$ are considered as foreground (green symbols) and those with redshift larger $ z_{\rm BCG1} + \sigma_z$ as background (orage symbols). We compute the average redshift for members of A1430-A and A1430-B and find a mean redshift difference $\Delta z = z_{\rm A} - z_{\rm B} = 0.003696$, or $\Delta v = 1109 \, \rm km \, s^{-1}$. If alternatively interpreted as a radial distance this would correspond to a separation of about $ 23\,\rm Mpc$.

  Do the two samples related to A1430-A and B represent two clusters, or are they drawn from the same galaxy population? There are several straightforward approaches to probe a distribution for unimodality versus bimodality, or multimodality. We used Sarle's bimodality coefficient
  \begin{equation}
    b 
    = 
    \frac{g^2 +1 }{k + 3x} 
    \quad \mbox{with} \quad 
    x = \frac{(N-1)^2}{(N-2)(N-3)}, 
  \end{equation}
  where $g$ is the sample skewness, $k$ is the sample excess kurtosis and $N$ is the number of objects in the sample \citep{Pfister_2013}.   A bimodality coefficient larger than the benchmark value $b_{\rm crit} = 5/9 \approx 0.555$ for a uniform distribution points towards bimodality or multimodality. For the galaxies in the two substructures, we find $b = 0.64$, indicating a bimodal distribution.

  The galaxies in the left part of Fig.\,\ref{fig::sdss_redshift} are scattered over the whole redshift range $z \approx 0.33 - 0.36$, which corresponds to a luminosity distance interval of $\sim 170\,\rm Mpc$ and might indicate a large-scale filament nearly perpendicular to the plane of the sky.

  The evidence for bimodality supports that the merger in A1430 does not take place in the plane of the sky but rather along the line of sight. The redshift difference of main cluster and subcluster indicates that either A1430-B is about 20\,Mpc less distant than A1430-A, is at same distance as A1430-A and moves with about $1000$\,km\,s$^{-1}$ towards the observer, or any combination of these two scenarios.   The velocity difference is typical for merging clusters. However, we stress that this relative velocity needs to be achieved with a separation in the plane of the sky of about 1\,Mpc.  The large scatter of galaxy redshifts may additionally support that we observe here a large cosmic filament along the line of sight.

% ===== SUBSEC :: simulated merger  ==================
% 
\subsection{A similar merger in a cosmological simulation}
\label{sec::simulation}

  % motivation, why we analyze a cosmological 
  To verify that it is plausible to assume that the subcluster moves with a significant velocity along the line of sight despite the separation of about 1\,Mpc in the plane of the sky, we inspect a large cosmological simulation in search for mergers with characteristics as close to A1430 as possible.
  
  % the 300 simu in a nutshell 
  The {\sc Three Hundred Project} \citep{2018MNRAS.480.2898C} introduces a large sample of hydrodynamical galaxy cluster re-simulations extracted from a periodic box of $1.43\,\rm Gpc$ comoving length on a side, including physics such as gas cooling, star formation, and stellar/AGN feedback. Mass resolution in zoomed-in regions corresponds to $m_{\rm DM}=1.29\times10^{9}\,$M$_\odot$ and $m_{\rm g}=2.71\times10^{8}\,$M$_\odot$ for dark-matter and gas particles, respectively.   The gravitational softening length for these particles in the high-resolution zones was set to $8.57\, \rm kpc $.   Some of the advantages of this sample include: an improved modelling of subgrid physics, and a state-of-the-art implementation of the smoothed-particle hydrodynamics (SPH) scheme. Moreover, the large number of available snapshots allows an excellent coverage of cluster evolution.   
  
  % selection criteria 
  We looked for a merging system with a similar mass ratio in a redshift range from 0.3 to 0.4.  However, since major mergers are rare, we focused on those with central-to-subcluster mass ratios below 3:1.   Similar to the clusters in A1430, we are mainly interested in separations of about $900\, \rm kpc$ in the plane of the sky and relative velocities along the line of sight of about $1100 \, \rm km \, s^{-1} $. 
  
  % final selection of example
  From our suite of simulated galaxy clusters, we selected a merging system at redshift 0.333 with a mass ratio of 5:1, being the mass of the main cluster $M_{500} = 7.4 \times 10^{14}\,$M$_\odot$.  Fig.~\ref{fig::sim_merger_1} shows the X-ray surface brightness (left panel) and flux-weighted gas temperature (right panel) for our merging cluster example. Since there is an offset between simulated and observed gas temperatures for the largest cluster masses (see, e.g., Fig.~9 of \citealt{2018MNRAS.480.2898C}), flux-weighted temperatures have been scaled to roughly match A1430's observed profile for the main cluster.   We note, however, that cluster simulations tend to have cooler cores than real ones.   While this renders a detailed comparison difficult, this does not affect our conclusions. 
  
  The projected distance between clusters in our example is $922\,$kpc; whereas their relative velocity along the line-of-sight is $1113\,$km\,s$^{-1}$, which is similar to the observed velocity difference of the subclusters in A1430. From the simulation snapshots, we observe that the subcluster is approaching, for the first time, towards the central one in a fairly radial orbit, still leading to an off-axis merger. Interestingly, as shown in the right panel of Fig.~\ref{fig::sim_merger_1}, the gas temperature increases between the clusters, indicating that the system is interacting.  Moreover, there is also a region with enhanced temperature {\it on a side} of the main cluster, resembling the cyan region in Fig.\,\ref{fig::XMM}.   In the simulation, this is produced by a minor substructure merging earlier with the central one.
  
  % conclusion
  This example illustrates that it is plausible to assume that the main and secondary clusters in A1430 are in the process of a merger. In principle, we cannot exclude the possibility that the observed redshift difference between the clusters could either indicate that A1430 actually consists of two still physically separated objects. However, the temperature increase observed in sectors 6 and 7 of A1430 most likely suggests that the two objects are already interacting. Moreover, the simulation shows that the observed redshift difference along the line-of-sight is not uncommon for merging galaxy clusters.

% ===== SUBSEC :: Conclusion  ==================
%  
\section{Conclusion}

We analysed the LOFAR, \textit{XMM-Newton}, \textit{Chandra}, VLA and SDSS data of the galaxy cluster Abell\,1430. The cluster undergoes a major merger and the LOTSS data reveal a very extended diffuse radio emission which is apparently comprised of a radio halo in the main cluster and a puzzling, extended radio emission related to the infalling subcluster. In detail, we found:

\begin{itemize}
    \item
      The X-ray surface brightness distribution reveals that A1430 is a double system, composed of a main cluster A1430-A and a subcluster A1430-B with a mass ratio 2:1. 
    \item 
      The LoTSS data reveal a radio halo for A1430-A. The morphology of the halo closely follows the X-ray surface brightness distribution of the main cluster component. The offset between the BCG, the X-ray concentration parameter and the centroid shift, and the missing cold core indicate that A1430-A is a disturbed system, which is typical for the clusters where a radio halo is present.
    \item 
      A large radio feature, which we dub `Pillow' owing to its morphology, is apparently associated with A1430-B. If indeed diffuse emission, this emission would be very unusual since it is located in a region of very low ICM density and it appears unlikely that it is a radio relic. We speculate that the two cluster components undergo an off-axis merger, providing the necessary energy dissipation for the Pillow radio feature. 
    \item
      The redshift distribution of galaxies in A1430, as obtained from spectroscopic information from SDSS, indicates that the two subclusters have a redshift difference of about $1100\,\rm km \, s^{-1}$. Together with the X-ray temperature distribution, this supports that A1430-B is indeed merging with A1430-A at a large impact parameter. We inspected a large-volume cosmological simulation and identified a merger with similar parameters that  shows a similar pattern of enhanced temperature in the interaction region. 
\end{itemize}

The LOFAR observations have revealed an atypical diffuse radio emission in a low-density environment, which put our understanding of the origin of diffuse radio emission onto test. 
\\

\begin{acknowledgements}
 
  This paper is based on data obtained with the International LOFAR Telescope (ILT) under project codes LC2\_038 and LC3\_008. LOFAR \citep{lofar_2013} is the LOw Frequency ARray designed and constructed by ASTRON. It has observing, data processing, and data storage facilities in several countries, which are owned by various parties (each with their own funding sources) and are collectively operated by the ILT foundation under a joint scientific policy. The ILT resources have benefited from the following recent major funding sources: CNRS-INSU, Observatoire de Paris and Université d’Orléans, France; BMBF, MIWF-NRW, MPG, Germany; Science Foundation Ireland (SFI), Department of Business, Enterprise and Innovation (DBEI), Ireland; NWO, The Netherlands; The Science and Technology Facilities Council, UK; Ministry of Science and Higher Education, Poland; The Istituto Nazionale di Astrofisica (INAF), Italy. This research made use of the Dutch national e-infrastructure with support of the SURF Cooperative (e-infra 180169) and the LOFAR e-infra group. The Jülich LOFAR Long Term Archive and the German LOFAR network are both coordinated and operated by the Jülich Supercomputing Centre (JSC). Computing resources on the supercomputer JUWELS at JSC were provided by the Gauss Centre for Supercomputing e.V. (grant CHTB00) through the John von Neumann Institute for Computing (NIC). This research made use of the University of Hertfordshire high-performance computing facility (http://uhhpc.herts.ac.uk) and the LOFAR-UK computing facility located at the University of Hertfordshire and supported by STFC [ST/P000096/1].
  \\
  
  CD acknowledges support by the German Academic Exchange Service (DAAD).
  AD acknowledges support by the BMBF Verbundforschung under the grant 05A20STA.
  KR acknowledges financial support from the ERC  Starting Grant ``MAGCOW", no. 714196.
  SEN acknowledges support by the Agencia Nacional de Promoci\'on Cient\'{\i}fica y Tecnol\'ogica (ANPCyT, PICT-2016-0667). He is member of the Carrera del Investigador Cient\'{\i}fico of CONICET. 
  RJvW and AB acknowledge support from the VIDI research programme with project number 639.042.729, which is financed by the Netherlands Organisation for Scientific Research (NWO).
  LL acknowledges financial contribution from the contracts ASI-INAF Athena 2019-27-HH.0, ``Attivit\`a di Studio per la comunit\`a scientifica di Astrofisica delle Alte Energie e Fisica Astroparticellare'' (Accordo Attuativo ASI-INAF n. 2017-14-H.0), and from INAF ``Call per interventi aggiuntivi a sostegno della ricerca di main stream di INA''.
  GY acknowledges financial support from MICIN/FEDER (Spain) under research grant PGC2018-094975-C21.
  \\
 
  The numerical simulations of the The Three Hundred collaboration used in this work  have been done at the MareNostrum Supercomputer of the BSC-CNS thanks to computing time granted by  The Red Española de Supercomputación and at the SuperMUC supercomputer of the `Leibniz-Rechenzentrum’ with CPU time assigned to the Project `pr83li’.
  \\
  
  This research has made use of data products from the Sloan Digital Sky Survey (SDSS). Funding for the SDSS and SDSS-II has been provided by the Alfred P. Sloan Foundation, the Participating Institutions (see below), the National Science Foundation, the National Aeronautics and Space Administration, the U.S. Department of Energy, the Japanese Monbukagakusho, the Max Planck Society, and the Higher Education Funding Council for England. The SDSS Web site is http://www.sdss.org/. The SDSS is managed by the Astrophysical Research Consortium (ARC) for the Participating Institutions. The Participating Institutions are: the American Museum of Natural History, Astrophysical Institute Potsdam, University of Basel, University of Cambridge (Cambridge University), Case Western Reserve University, the University of Chicago, the Fermi National Accelerator Laboratory (Fermilab), the Institute for Advanced Study, the Japan Participation Group, the Johns Hopkins University, the Joint Institute for Nuclear Astrophysics, the Kavli Institute for Particle Astrophysics and Cosmology, the Korean Scientist Group, the Los Alamos National Laboratory, the Max-Planck-Institute for Astronomy (MPIA), the Max-Planck-Institute for Astrophysics (MPA), the New Mexico State University, the Ohio State University, the University of Pittsburgh, University of Portsmouth, Princeton University, the United States Naval Observatory, and the University of Washington. 
  
\end{acknowledgements}

%  \begin{figure*}[htbp]
%      \centering
%      \includegraphics[scale=0.3]{A1430_LOFAR-lores_XMM.png}
%      \caption{\textbf{A1430}. \textit{Colorscale:} XMM-Newton image. \textit{Black contours:} Total power radio map of the LOFAR HBA observation. Noise level is ... Contours start at $3\sigma_{\text{rms}}$ and are spaced by $\sqrt{2}$. \textit{Magenta ellipse:} The restoring beam size is $35^{\prime\prime}\times27^{\prime\prime}$ with a position angle of $-8^{\circ}$. \textit{Blue crosses:} Location of galaxies associated to A1430. \textit{Blue circles:} Associated BCGs of A1430.}
%      \label{fig:A1430_XMM}
%  \end{figure*}{}

\bibliographystyle{aa}
\bibliography{refs}

\begin{thebibliography}{68}
\expandafter\ifx\csname natexlab\endcsname\relax\def\natexlab#1{#1}\fi

\bibitem[{{Abell}(1958)}]{abell_1958}
{Abell}, G.~O. 1958, \apjs, 3, 211

\bibitem[{{Aguado} {et~al.}(2019){Aguado}, {Ahumada}, {Almeida}, {Anderson},
  {Andrews}, {Anguiano}, {Aquino Ort{\'\i}z}, {Arag{\'o}n-Salamanca},
  {Argudo-Fern{\'a}ndez}, {Aubert}, {Avila-Reese}, {Badenes}, {Barboza
  Rembold}, {Barger}, {Barrera-Ballesteros}, {Bates}, {Bautista}, {Beaton},
  {Beers}, {Belfiore}, {Bernardi}, {Bershady}, {Beutler}, {Bird}, {Bizyaev},
  {Blanc}, {Blanton}, {Blomqvist}, {Bolton}, {Boquien}, {Borissova}, {Bovy},
  {Brand t}, {Brinkmann}, {Brownstein}, {Bundy}, {Burgasser}, {Byler}, {Cano
  Diaz}, {Cappellari}, {Carrera}, {Cervantes Sodi}, {Chen}, {Cherinka}, {Choi},
  {Chung}, {Coffey}, {Comerford}, {Comparat}, {Covey}, {da Silva Ilha}, {da
  Costa}, {Dai}, {Damke}, {Darling}, {Davies}, {Dawson}, {de Sainte Agathe},
  {Deconto Machado}, {Del Moro}, {De Lee}, {Diamond-Stanic}, {Dom{\'\i}nguez
  S{\'a}nchez}, {Donor}, {Drory}, {du Mas des Bourboux}, {Duckworth}, {Dwelly},
  {Ebelke}, {Emsellem}, {Escoffier}, {Fern{\'a}ndez-Trincado}, {Feuillet},
  {Fischer}, {Fleming}, {Fraser-McKelvie}, {Freischlad}, {Frinchaboy}, {Fu},
  {Galbany}, {Garcia-Dias}, {Garc{\'\i}a-Hern{\'a}ndez}, {Garma Oehmichen},
  {Geimba Maia}, {Gil-Mar{\'\i}n}, {Grabowski}, {Gu}, {Guo}, {Ha},
  {Harrington}, {Hasselquist}, {Hayes}, {Hearty}, {Hernandez Toledo}, {Hicks},
  {Hogg}, {Holley-Bockelmann}, {Holtzman}, {Hsieh}, {Hunt}, {Hwang},
  {Ibarra-Medel}, {Jimenez Angel}, {Johnson}, {Jones}, {J{\"o}nsson},
  {Kinemuchi}, {Kollmeier}, {Krawczyk}, {Kreckel}, {Kruk}, {Lacerna}, {Lan},
  {Lane}, {Law}, {Lee}, {Li}, {Lian}, {Lin}, {Lin}, {Lintott}, {Long},
  {Longa-Pe{\~n}a}, {Mackereth}, {de la Macorra}, {Majewski}, {Malanushenko},
  {Manchado}, {Maraston}, {Mariappan}, {Marinelli}, {Marques-Chaves},
  {Masseron}, {Masters}, {McDermid}, {Medina Pe{\~n}a}, {Meneses-Goytia},
  {Merloni}, {Merrifield}, {Meszaros}, {Minniti}, {Minsley}, {Muna}, {Myers},
  {Nair}, {Correa do Nascimento}, {Newman}, {Nitschelm}, {Olmstead}, {Oravetz},
  {Oravetz}, {Ortega Minakata}, {Pace}, {Padilla}, {Palicio}, {Pan}, {Pan},
  {Parikh}, {Parker}, {Peirani}, {Penny}, {Percival}, {Perez-Fournon},
  {Peterken}, {Pinsonneault}, {Prakash}, {Raddick}, {Raichoor}, {Riffel},
  {Riffel}, {Rix}, {Robin}, {Roman-Lopes}, {Rose}, {Ross}, {Rossi}, {Rowlands},
  {Rubin}, {S{\'a}nchez}, {S{\'a}nchez-Gallego}, {Sayres}, {Schaefer},
  {Schiavon}, {Schimoia}, {Schlafly}, {Schlegel}, {Schneider}, {Schultheis},
  {Seo}, {Shamsi}, {Shao}, {Shen}, {Shetty}, {Simonian}, {Smethurst}, {Sobeck},
  {Souter}, {Spindler}, {Stark}, {Stassun}, {Steinmetz}, {Storchi-Bergmann},
  {Stringfellow}, {Su{\'a}rez}, {Sun}, {Taghizadeh-Popp}, {Talbot}, {Tayar},
  {Thakar}, {Thomas}, {Tissera}, {Tojeiro}, {Troup}, {Unda-Sanzana},
  {Valenzuela}, {Vargas-Maga{\~n}a}, {V{\'a}zquez-Mata}, {Wake}, {Weaver},
  {Weijmans}, {Westfall}, {Wild}, {Wilson}, {Woods}, {Yan}, {Yang}, {Zamora},
  {Zasowski}, {Zhang}, {Zheng}, {Zheng}, {Zhu}, {Zinn}, \&
  {Zou}}]{2019ApJS..240...23A}
{Aguado}, D.~S., {Ahumada}, R., {Almeida}, A., {et~al.} 2019, \apjs, 240, 23

\bibitem[{{Akamatsu} \& {Kawahara}(2013)}]{2013PASJ...65...16A}
{Akamatsu}, H. \& {Kawahara}, H. 2013, \pasj, 65, 16

\bibitem[{{Bacchi} {et~al.}(2003){Bacchi}, {Feretti}, {Giovannini}, \&
  {Govoni}}]{2003A&A...400..465B}
{Bacchi}, M., {Feretti}, L., {Giovannini}, G., \& {Govoni}, F. 2003, \aap, 400,
  465

\bibitem[{{B{\"o}hringer} {et~al.}(2013){B{\"o}hringer}, {Chon}, {Collins},
  {Guzzo}, {Nowak}, \& {Bobrovskyi}}]{2013A&A...555A..30B}
{B{\"o}hringer}, H., {Chon}, G., {Collins}, C.~A., {et~al.} 2013, \aap, 555,
  A30

\bibitem[{{B{\"o}hringer} {et~al.}(2017){B{\"o}hringer}, {Chon}, {Retzlaff},
  {Tr{\"u}mper}, {Meisenheimer}, \& {Schartel}}]{2017AJ....153..220B}
{B{\"o}hringer}, H., {Chon}, G., {Retzlaff}, J., {et~al.} 2017, \aj, 153, 220

\bibitem[{{B{\"o}hringer} {et~al.}(2000){B{\"o}hringer}, {Voges}, {Huchra},
  {McLean}, {Giacconi}, {Rosati}, {Burg}, {Mader}, {Schuecker}, {Simi{\c c}},
  {Komossa}, {Reiprich}, {Retzlaff}, \& {Tr{\"u}mper}}]{noras_2000}
{B{\"o}hringer}, H., {Voges}, W., {Huchra}, J.~P., {et~al.} 2000, \apjs, 129,
  435

\bibitem[{{Bonafede} {et~al.}(2009){Bonafede}, {Giovannini}, {Feretti},
  {Govoni}, \& {Murgia}}]{2009A&A...494..429B}
{Bonafede}, A., {Giovannini}, G., {Feretti}, L., {Govoni}, F., \& {Murgia}, M.
  2009, \aap, 494, 429

\bibitem[{{Botteon} {et~al.}(2016){Botteon}, {Gastaldello}, {Brunetti}, \&
  {Kale}}]{2016MNRAS.463.1534B}
{Botteon}, A., {Gastaldello}, F., {Brunetti}, G., \& {Kale}, R. 2016, \mnras,
  463, 1534

\bibitem[{{Botteon} {et~al.}(2018{\natexlab{a}}){Botteon}, {Shimwell},
  {Bonafede}, {Dallacasa}, {Brunetti}, {Mandal}, {van Weeren}, {Br{\"u}ggen},
  {Cassano}, {de Gasperin}, {Hoang}, {Hoeft}, {R{\"o}ttgering}, {Savini},
  {White}, {Wilber}, \& {Venturi}}]{botteon_2018}
{Botteon}, A., {Shimwell}, T.~W., {Bonafede}, A., {et~al.} 2018{\natexlab{a}},
  \mnras, 478, 885

\bibitem[{{Botteon} {et~al.}(2018{\natexlab{b}}){Botteon}, {Shimwell},
  {Bonafede}, {Dallacasa}, {Brunetti}, {Mandal}, {van Weeren}, {Br{\"u}ggen},
  {Cassano}, {de Gasperin}, {Hoang}, {Hoeft}, {R{\"o}ttgering}, {Savini},
  {White}, {Wilber}, \& {Venturi}}]{2018MNRAS.478..885B}
{Botteon}, A., {Shimwell}, T.~W., {Bonafede}, A., {et~al.} 2018{\natexlab{b}},
  \mnras, 478, 885

\bibitem[{{Botteon} {et~al.}(2020){Botteon}, {van Weeren}, {Brunetti}, {de
  Gasperin}, {Intema}, {Osinga}, {Di Gennaro}, {Shimwell}, {Bonafede},
  {Br{\"u}ggen}, {Cassano}, {Cuciti}, {Dallacasa}, {Gastaldello}, {Mandal},
  {Rossetti}, \& {R{\"o}ttgering}}]{2020MNRAS.499L..11B}
{Botteon}, A., {van Weeren}, R.~J., {Brunetti}, G., {et~al.} 2020, \mnras, 499,
  L11

\bibitem[{{Bourdin} {et~al.}(2013){Bourdin}, {Mazzotta}, {Markevitch},
  {Giacintucci}, \& {Brunetti}}]{2013ApJ...764...82B}
{Bourdin}, H., {Mazzotta}, P., {Markevitch}, M., {Giacintucci}, S., \&
  {Brunetti}, G. 2013, \apj, 764, 82

\bibitem[{{Briggs}(1995)}]{1995AAS...18711202B}
{Briggs}, D.~S. 1995, in American Astronomical Society Meeting Abstracts, Vol.
  187, American Astronomical Society Meeting Abstracts, 112.02

\bibitem[{{Brown} {et~al.}(2011){Brown}, {Emerick}, {Rudnick}, \&
  {Brunetti}}]{2011ApJ...740L..28B}
{Brown}, S., {Emerick}, A., {Rudnick}, L., \& {Brunetti}, G. 2011, \apjl, 740,
  L28

\bibitem[{{Brunetti} \& {Jones}(2014)}]{brunetti_2014}
{Brunetti}, G. \& {Jones}, T.~W. 2014, International Journal of Modern Physics
  D, 23, 1430007

\bibitem[{{Brunetti} \& {Vazza}(2020)}]{2020PhRvL.124e1101B}
{Brunetti}, G. \& {Vazza}, F. 2020, \prl, 124, 051101

\bibitem[{{Buote}(2001)}]{buote_2001}
{Buote}, D.~A. 2001, \apjl, 553, L15

\bibitem[{{Cassano} {et~al.}(2013){Cassano}, {Ettori}, {Brunetti},
  {Giacintucci}, {Pratt}, {Venturi}, {Kale}, {Dolag}, \&
  {Markevitch}}]{2013ApJ...777..141C}
{Cassano}, R., {Ettori}, S., {Brunetti}, G., {et~al.} 2013, \apj, 777, 141

\bibitem[{{Cassano} {et~al.}(2010){Cassano}, {Ettori}, {Giacintucci},
  {Brunetti}, {Markevitch}, {Venturi}, \& {Gitti}}]{2010ApJ...721L..82C}
{Cassano}, R., {Ettori}, S., {Giacintucci}, S., {et~al.} 2010, \apjl, 721, L82

\bibitem[{{Clarke} \& {Ensslin}(2006)}]{2006AJ....131.2900C}
{Clarke}, T.~E. \& {Ensslin}, T.~A. 2006, \aj, 131, 2900

\bibitem[{{Cuciti} {et~al.}(2018){Cuciti}, {Brunetti}, {van Weeren},
  {Bonafede}, {Dallacasa}, {Cassano}, {Venturi}, \&
  {Kale}}]{2018A&A...609A..61C}
{Cuciti}, V., {Brunetti}, G., {van Weeren}, R., {et~al.} 2018, \aap, 609, A61

\bibitem[{{Cui} {et~al.}(2018){Cui}, {Knebe}, {Yepes}, {Pearce}, {Power},
  {Dave}, {Arth}, {Borgani}, {Dolag}, {Elahi}, {Mostoghiu}, {Murante}, {Rasia},
  {Stoppacher}, {Vega-Ferrero}, {Wang}, {Yang}, {Benson}, {Cora}, {Croton},
  {Sinha}, {Stevens}, {Vega-Mart{\'\i}nez}, {Arthur}, {Baldi}, {Ca{\~n}as},
  {Cialone}, {Cunnama}, {De Petris}, {Durando}, {Ettori}, {Gottl{\"o}ber},
  {Nuza}, {Old}, {Pilipenko}, {Sorce}, \& {Welker}}]{2018MNRAS.480.2898C}
{Cui}, W., {Knebe}, A., {Yepes}, G., {et~al.} 2018, \mnras, 480, 2898

\bibitem[{{Ebeling} {et~al.}(2000){Ebeling}, {Edge}, \& {Henry}}]{ebeling_2000}
{Ebeling}, H., {Edge}, A.~C., \& {Henry}, J.~P. 2000, in Bulletin of the
  American Astronomical Society, Vol.~32, AAS/High Energy Astrophysics Division
  \#5, 1209

\bibitem[{{Eckert} {et~al.}(2017){Eckert}, {Ettori}, {Pointecouteau},
  {Molendi}, {Paltani}, \& {Tchernin}}]{eckert_2017}
{Eckert}, D., {Ettori}, S., {Pointecouteau}, E., {et~al.} 2017, Astronomische
  Nachrichten, 338, 293

\bibitem[{{Feretti} {et~al.}(2012){Feretti}, {Giovannini}, {Govoni}, \&
  {Murgia}}]{feretti_2012}
{Feretti}, L., {Giovannini}, G., {Govoni}, F., \& {Murgia}, M. 2012, \aapr, 20,
  54

\bibitem[{{Finoguenov} {et~al.}(2010){Finoguenov}, {Sarazin}, {Nakazawa},
  {Wik}, \& {Clarke}}]{2010ApJ...715.1143F}
{Finoguenov}, A., {Sarazin}, C.~L., {Nakazawa}, K., {Wik}, D.~R., \& {Clarke},
  T.~E. 2010, \apj, 715, 1143

\bibitem[{{Ghirardini} {et~al.}(2019){Ghirardini}, {Eckert}, {Ettori},
  {Pointecouteau}, {Molendi}, {Gaspari}, {Rossetti}, {De Grandi}, {Roncarelli},
  {Bourdin}, {Mazzotta}, {Rasia}, \& {Vazza}}]{ghirardini_2019}
{Ghirardini}, V., {Eckert}, D., {Ettori}, S., {et~al.} 2019, \aap, 621, A41

\bibitem[{{Giovannini} {et~al.}(2009){Giovannini}, {Bonafede}, {Feretti},
  {Govoni}, {Murgia}, {Ferrari}, \& {Monti}}]{2009A&A...507.1257G}
{Giovannini}, G., {Bonafede}, A., {Feretti}, L., {et~al.} 2009, \aap, 507, 1257

\bibitem[{{Golovich} {et~al.}(2019){Golovich}, {Dawson}, {Wittman}, {van
  Weeren}, {Andrade-Santos}, {Jee}, {Benson}, {de Gasperin}, {Venturi},
  {Bonafede}, {Sobral}, {Ogrean}, {Lemaux}, {Brada{\v{c}}}, {Br{\"u}ggen}, \&
  {Peter}}]{2019ApJ...882...69G}
{Golovich}, N., {Dawson}, W.~A., {Wittman}, D.~M., {et~al.} 2019, \apj, 882, 69

\bibitem[{{Govoni} {et~al.}(2001){Govoni}, {Feretti}, {Giovannini},
  {B{\"o}hringer}, {Reiprich}, \& {Murgia}}]{2001A&A...376..803G}
{Govoni}, F., {Feretti}, L., {Giovannini}, G., {et~al.} 2001, \aap, 376, 803

\bibitem[{{Govoni} {et~al.}(2019){Govoni}, {Orr{\`u}}, {Bonafede}, {Iacobelli},
  {Paladino}, {Vazza}, {Murgia}, {Vacca}, {Giovannini}, {Feretti}, {Loi},
  {Bernardi}, {Ferrari}, {Pizzo}, {Gheller}, {Manti}, {Br{\"u}ggen},
  {Brunetti}, {Cassano}, {de Gasperin}, {En{\ss}lin}, {Hoeft}, {Horellou},
  {Junklewitz}, {R{\"o}ttgering}, {Scaife}, {Shimwell}, {van Weeren}, \&
  {Wise}}]{2019Sci...364..981G}
{Govoni}, F., {Orr{\`u}}, E., {Bonafede}, A., {et~al.} 2019, Science, 364, 981

\bibitem[{{Hao} {et~al.}(2010){Hao}, {McKay}, {Koester}, {Rykoff}, {Rozo},
  {Annis}, {Wechsler}, {Evrard}, {Siegel}, {Becker}, {Busha}, {Gerdes},
  {Johnston}, \& {Sheldon}}]{2010ApJS..191..254H}
{Hao}, J., {McKay}, T.~A., {Koester}, B.~P., {et~al.} 2010, \apjs, 191, 254

\bibitem[{{Hoang} {et~al.}(2018){Hoang}, {Shimwell}, {van Weeren}, {Intema},
  {R{\"o}ttgering}, {Andrade-Santos}, {Akamatsu}, {Bonafede}, {Brunetti},
  {Dawson}, {Golovich}, {Best}, {Botteon}, {Br{\"u}ggen}, {Cassano}, {de
  Gasperin}, {Hoeft}, {Stroe}, \& {White}}]{2018MNRAS.478.2218H}
{Hoang}, D.~N., {Shimwell}, T.~W., {van Weeren}, R.~J., {et~al.} 2018, \mnras,
  478, 2218

\bibitem[{{Kierdorf} {et~al.}(2017){Kierdorf}, {Beck}, {Hoeft}, {Klein}, {van
  Weeren}, {Forman}, \& {Jones}}]{2017A&A...600A..18K}
{Kierdorf}, M., {Beck}, R., {Hoeft}, M., {et~al.} 2017, \aap, 600, A18

\bibitem[{{Markevitch} {et~al.}(2002){Markevitch}, {Gonzalez}, {David},
  {Vikhlinin}, {Murray}, {Forman}, {Jones}, \& {Tucker}}]{2002ApJ...567L..27M}
{Markevitch}, M., {Gonzalez}, A.~H., {David}, L., {et~al.} 2002, \apjl, 567,
  L27

\bibitem[{{Offringa} {et~al.}(2010){Offringa}, {de Bruyn}, {Biehl}, {Zaroubi},
  {Bernardi}, \& {Pandey}}]{Offringa2010}
{Offringa}, A.~R., {de Bruyn}, A.~G., {Biehl}, M., {et~al.} 2010, \mnras, 405,
  155

\bibitem[{Offringa {et~al.}(2014)Offringa, McKinley, Hurley-Walker,
  {et~al.}}]{offringa-wsclean-2014}
Offringa, A.~R., McKinley, B., Hurley-Walker, {et~al.} 2014, MNRAS, 444, 606

\bibitem[{{Owen} {et~al.}(2014){Owen}, {Rudnick}, {Eilek}, {Rau}, {Bhatnagar},
  \& {Kogan}}]{2014ApJ...794...24O}
{Owen}, F.~N., {Rudnick}, L., {Eilek}, J., {et~al.} 2014, \apj, 794, 24

\bibitem[{{Perley} \& {Butler}(2013)}]{Perley2013}
{Perley}, R.~A. \& {Butler}, B.~J. 2013, \apjs, 204, 19

\bibitem[{Pfister {et~al.}(2013)Pfister, Schwarz, Janczyk, Dale, \&
  Freeman}]{Pfister_2013}
Pfister, R., Schwarz, K., Janczyk, M., Dale, R., \& Freeman, J. 2013, Frontiers
  in Psychology, 4

\bibitem[{{Planck Collaboration} {et~al.}(2015){Planck Collaboration}, {Ade},
  {Aghanim}, {Armitage-Caplan}, {Arnaud}, {Ashdown}, {Atrio-Barand ela},
  {Aumont}, {Aussel}, {Baccigalupi}, {Band ay}, {Barreiro}, {Barrena},
  {Bartelmann}, {Bartlett}, {Battaner}, {Benabed}, {Beno{\^\i}t},
  {Benoit-L{\'e}vy}, {Bernard}, {Bersanelli}, {Bielewicz}, {Bikmaev}, {Bobin},
  {Bock}, {B{\"o}hringer}, {Bonaldi}, {Bond}, {Borrill}, {Bouchet}, {Bridges},
  {Bucher}, {Burenin}, {Burigana}, {Butler}, {Cardoso}, {Carvalho}, {Catalano},
  {Challinor}, {Chamballu}, {Chary}, {Chen}, {Chiang}, {Chiang}, {Chon},
  {Christensen}, {Churazov}, {Church}, {Clements}, {Colombi}, {Colombo},
  {Comis}, {Couchot}, {Coulais}, {Crill}, {Curto}, {Cuttaia}, {Da Silva},
  {Dahle}, {Danese}, {Davies}, {Davis}, {de Bernardis}, {de Rosa}, {de Zotti},
  {Delabrouille}, {Delouis}, {D{\'e}mocl{\`e}s}, {D{\'e}sert}, {Dickinson},
  {Diego}, {Dolag}, {Dole}, {Donzelli}, {Dor{\'e}}, {Douspis}, {Dupac},
  {Efstathiou}, {En{\ss}lin}, {Eriksen}, {Feroz}, {Ferragamo}, {Finelli},
  {Flores-Cacho}, {Forni}, {Frailis}, {Franceschi}, {Fromenteau}, {Galeotta},
  {Ganga}, {G{\'e}nova-Santos}, {Giard}, {Giardino}, {Gilfanov},
  {Giraud-H{\'e}raud}, {Gonz{\'a}lez-Nuevo}, {G{\'o}rski}, {Grainge},
  {Gratton}, {Gregorio}, {Groeneboom}, {Gruppuso}, {Hansen}, {Hanson},
  {Harrison}, {Hempel}, {Henrot-Versill{\'e}}, {Hern{\'a}ndez-Monteagudo},
  {Herranz}, {Hildebrandt}, {Hivon}, {Hobson}, {Holmes}, {Hornstrup}, {Hovest},
  {Huffenberger}, {Hurier}, {Hurley-Walker}, {Jaffe}, {Jaffe}, {Jones},
  {Juvela}, {Keih{\"a}nen}, {Keskitalo}, {Khamitov}, {Kisner}, {Kneissl},
  {Knoche}, {Knox}, {Kunz}, {Kurki-Suonio}, {Lagache}, {L{\"a}hteenm{\"a}ki},
  {Lamarre}, {Lasenby}, {Laureijs}, {Lawrence}, {Leahy}, {Leonardi},
  {Le{\'o}n-Tavares}, {Lesgourgues}, {Li}, {Liddle}, {Liguori}, {Lilje},
  {Linden-V{\o}rnle}, {L{\'o}pez-Caniego}, {Lubin}, {Mac{\'\i}as-P{\'e}rez},
  {MacTavish}, {Maffei}, {Maino}, {Mandolesi}, {Maris}, {Marshall}, {Martin},
  {Mart{\'\i}nez-Gonz{\'a}lez}, {Masi}, {Massardi}, {Matarrese}, {Matthai},
  {Mazzotta}, {Mei}, {Meinhold}, {Melchiorri}, {Melin}, {Mendes}, {Mennella},
  {Migliaccio}, {Mikkelsen}, {Mitra}, {Miville-Desch{\^e}nes}, {Moneti},
  {Montier}, {Morgante}, {Mortlock}, {Munshi}, {Murphy}, {Naselsky}, {Nastasi},
  {Nati}, {Natoli}, {Nesvadba}, {Netterfield}, {N{\o}rgaard-Nielsen},
  {Noviello}, {Novikov}, {Novikov}, {O'Dwyer}, {Olamaie}, {Osborne},
  {Oxborrow}, {Paci}, {Pagano}, {Pajot}, {Paoletti}, {Pasian}, {Patanchon},
  {Pearson}, {Perdereau}, {Perotto}, {Perrott}, {Perrotta}, {Piacentini},
  {Piat}, {Pierpaoli}, {Pietrobon}, {Plaszczynski}, {Pointecouteau}, {Polenta},
  {Ponthieu}, {Popa}, {Poutanen}, {Pratt}, {Pr{\'e}zeau}, {Prunet}, {Puget},
  {Rachen}, {Reach}, {Rebolo}, {Reinecke}, {Remazeilles}, {Renault},
  {Ricciardi}, {Riller}, {Ristorcelli}, {Rocha}, {Rosset}, {Roudier},
  {Rowan-Robinson}, {Rubi{\~n}o-Mart{\'\i}n}, {Rumsey}, {Rusholme}, {Sandri},
  {Santos}, {Saunders}, {Savini}, {Schammel}, {Scott}, {Seiffert}, {Shellard},
  {Shimwell}, {Spencer}, {Starck}, {Stolyarov}, {Stompor}, {Streblyanska},
  {Sudiwala}, {Sunyaev}, {Sureau}, {Sutton}, {Suur-Uski}, {Sygnet}, {Tauber},
  {Tavagnacco}, {Terenzi}, {Toffolatti}, {Tomasi}, {Tramonte}, {Tristram},
  {Tucci}, {Tuovinen}, {T{\"u}rler}, {Umana}, {Valenziano}, {Valiviita}, {Van
  Tent}, {Vibert}, {Vielva}, {Villa}, {Vittorio}, {Wade}, {Wandelt}, {White},
  {White}, {Yvon}, {Zacchei}, \& {Zonca}}]{2015A&A...581A..14P}
{Planck Collaboration}, {Ade}, P.~A.~R., {Aghanim}, N., {et~al.} 2015, \aap,
  581, A14

\bibitem[{{Planck Collaboration} {et~al.}(2011){Planck Collaboration}, {Ade},
  {Aghanim}, {Arnaud}, {Ashdown}, {Aumont}, {Baccigalupi}, {Baker}, {Balbi},
  {Banday}, {Barreiro}, {Bartlett}, {Battaner}, {Benabed}, {Bennett},
  {Beno{\^\i}t}, {Bernard}, {Bersanelli}, {Bhatia}, {Bock}, {Bonaldi}, {Bond},
  {Borrill}, {Bouchet}, {Bradshaw}, {Bremer}, {Bucher}, {Burigana}, {Butler},
  {Cabella}, {Cantalupo}, {Cappellini}, {Cardoso}, {Carr}, {Casale},
  {Catalano}, {Cay{\'o}n}, {Challinor}, {Chamballu}, {Charra}, {Chary},
  {Chiang}, {Chiang}, {Christensen}, {Clements}, {Colombi}, {Couchot},
  {Coulais}, {Crill}, {Crone}, {Crook}, {Cuttaia}, {Danese}, {D'Arcangelo},
  {Davies}, {Davis}, {de Bernardis}, {de Bruin}, {de Gasperis}, {de Rosa}, {de
  Zotti}, {Delabrouille}, {Delouis}, {D{\'e}sert}, {Dick}, {Dickinson},
  {Dolag}, {Dole}, {Donzelli}, {Dor{\'e}}, {D{\"o}rl}, {Douspis}, {Dupac},
  {Efstathiou}, {En{\ss}lin}, {Eriksen}, {Finelli}, {Foley}, {Forni},
  {Fosalba}, {Frailis}, {Franceschi}, {Freschi}, {Gaier}, {Galeotta},
  {Gallegos}, {Gandolfo}, {Ganga}, {Giard}, {Giardino}, {Gienger},
  {Giraud-H{\'e}raud}, {Gonz{\'a}lez}, {Gonz{\'a}lez-Nuevo}, {G{\'o}rski},
  {Gratton}, {Gregorio}, {Gruppuso}, {Guyot}, {Haissinski}, {Hansen},
  {Harrison}, {Helou}, {Henrot- Versill{\'e}}, {Hern{\'a}ndez-Monteagudo},
  {Herranz}, {Hildebrandt}, {Hivon}, {Hobson}, {Holmes}, {Hornstrup}, {Hovest},
  {Hoyland}, {Huffenberger}, {Jaffe}, {Jagemann}, {Jones}, {Juillet}, {Juvela},
  {Kangaslahti}, {Keih{\"a}nen}, {Keskitalo}, {Kisner}, {Kneissl}, {Knox},
  {Krassenburg}, {Kurki- Suonio}, {Lagache}, {L{\"a}hteenm{\"a}ki}, {Lamarre},
  {Lange}, {Lasenby}, {Laureijs}, {Lawrence}, {Leach}, {Leahy}, {Leonardi},
  {Leroy}, {Lilje}, {Linden-V{\o}rnle}, {L{\'o}pez-Caniego}, {Lowe}, {Lubin},
  {Mac{\'\i}as-P{\'e}rez}, {Maciaszek}, {MacTavish}, {Maffei}, {Maino},
  {Mandolesi}, {Mann}, {Maris}, {Mart{\'\i}nez-Gonz{\'a}lez}, {Masi},
  {Massardi}, {Matarrese}, {Matthai}, {Mazzotta}, {McDonald}, {McGehee},
  {Meinhold}, {Melchiorri}, {Melin}, {Mendes}, {Mennella}, {Mevi},
  {Miniscalco}, {Mitra}, {Miville-Desch{\^e}nes}, {Moneti}, {Montier},
  {Morgante}, {Morisset}, {Mortlock}, {Munshi}, {Murphy}, {Naselsky}, {Natoli},
  {Netterfield}, {N{\o}rgaard-Nielsen}, {Noviello}, {Novikov}, {Novikov},
  {O'Dwyer}, {Ortiz}, {Osborne}, {Osuna}, {Oxborrow}, {Pajot}, {Paladini},
  {Partridge}, {Pasian}, {Passvogel}, {Patanchon}, {Pearson}, {Pearson},
  {Perdereau}, {Perotto}, {Perrotta}, {Piacentini}, {Piat}, {Pierpaoli},
  {Plaszczynski}, {Platania}, {Pointecouteau}, {Polenta}, {Ponthieu}, {Popa},
  {Poutanen}, {Pr{\'e}zeau}, {Prunet}, {Puget}, {Rachen}, {Reach}, {Rebolo},
  {Reinecke}, {Reix}, {Renault}, {Ricciardi}, {Riller}, {Ristorcelli}, {Rocha},
  {Rosset}, {Rowan-Robinson}, {Rubi{\~n}o-Mart{\'\i}n}, {Rusholme}, {Salerno},
  {Sandri}, {Santos}, {Savini}, {Schaefer}, {Scott}, {Seiffert}, {Shellard},
  {Simonetto}, {Smoot}, {Sozzi}, {Starck}, {Sternberg}, {Stivoli}, {Stolyarov},
  {Stompor}, {Stringhetti}, {Sudiwala}, {Sunyaev}, {Sygnet}, {Tapiador},
  {Tauber}, {Tavagnacco}, {Taylor}, {Terenzi}, {Texier}, {Toffolatti},
  {Tomasi}, {Torre}, {Tristram}, {Tuovinen}, {T{\"u}rler}, {Tuttlebee},
  {Umana}, {Valenziano}, {Valiviita}, {Varis}, {Vibert}, {Vielva}, {Villa},
  {Vittorio}, {Wade}, {Wandelt}, {Watson}, {White}, {White}, {Wilkinson},
  {Yvon}, {Zacchei}, \& {Zonca}}]{planck_2011}
{Planck Collaboration}, {Ade}, P.~A.~R., {Aghanim}, N., {et~al.} 2011, \aap,
  536, A1

\bibitem[{{Planck Collaboration} {et~al.}(2016){Planck Collaboration}, {Ade},
  {Aghanim}, {Arnaud}, {Ashdown}, {Aumont}, {Baccigalupi}, {Banday},
  {Barreiro}, \& {Barrena}}]{2016A&A...594A..27P}
{Planck Collaboration}, {Ade}, P.~A.~R., {Aghanim}, N., {et~al.} 2016, \aap,
  594, A27

\bibitem[{{Pratt} {et~al.}(2009){Pratt}, {Croston}, {Arnaud}, \&
  {B{\"o}hringer}}]{2009A&A...498..361P}
{Pratt}, G.~W., {Croston}, J.~H., {Arnaud}, M., \& {B{\"o}hringer}, H. 2009,
  \aap, 498, 361

\bibitem[{{Rossetti} {et~al.}(2016){Rossetti}, {Gastaldello}, {Ferioli},
  {Bersanelli}, {De Grandi}, {Eckert}, {Ghizzardi}, {Maino}, \&
  {Molendi}}]{2016MNRAS.457.4515R}
{Rossetti}, M., {Gastaldello}, F., {Ferioli}, G., {et~al.} 2016, \mnras, 457,
  4515

\bibitem[{{Rozo} {et~al.}(2015){Rozo}, {Rykoff}, {Bartlett}, \&
  {Melin}}]{rozo_2015}
{Rozo}, E., {Rykoff}, E.~S., {Bartlett}, J.~G., \& {Melin}, J.-B. 2015, \mnras,
  450, 592

\bibitem[{{Sanderson} {et~al.}(2009){Sanderson}, {Edge}, \&
  {Smith}}]{2009MNRAS.398.1698S}
{Sanderson}, A. J.~R., {Edge}, A.~C., \& {Smith}, G.~P. 2009, \mnras, 398, 1698

\bibitem[{{Shimwell} {et~al.}(2019){Shimwell}, {Tasse}, {Hardcastle}, {Mechev},
  {Williams}, {Best}, {R{\"o}ttgering}, {Callingham}, {Dijkema}, {de Gasperin},
  {Hoang}, {Hugo}, {Mirmont}, {Oonk}, {Prandoni}, {Rafferty}, {Sabater},
  {Smirnov}, {van Weeren}, {White}, {Atemkeng}, {Bester}, {Bonnassieux},
  {Br{\"u}ggen}, {Brunetti}, {Chy{\.z}y}, {Cochrane}, {Conway}, {Croston},
  {Danezi}, {Duncan}, {Haverkorn}, {Heald}, {Iacobelli}, {Intema}, {Jackson},
  {Jamrozy}, {Jarvis}, {Lakhoo}, {Mevius}, {Miley}, {Morabito}, {Morganti},
  {Nisbet}, {Orr{\'u}}, {Perkins}, {Pizzo}, {Schrijvers}, {Smith}, {Vermeulen},
  {Wise}, {Alegre}, {Bacon}, {van Bemmel}, {Beswick}, {Bonafede}, {Botteon},
  {Bourke}, {Brienza}, {Calistro Rivera}, {Cassano}, {Clarke}, {Conselice},
  {Dettmar}, {Drabent}, {Dumba}, {Emig}, {En{\ss}lin}, {Ferrari}, {Garrett},
  {G{\'e}nova-Santos}, {Goyal}, {G{\"u}rkan}, {Hale}, {Harwood}, {Heesen},
  {Hoeft}, {Horellou}, {Jackson}, {Kokotanekov}, {Kondapally},
  {Kunert-Bajraszewska}, {Mahatma}, {Mahony}, {Mandal}, {McKean}, {Merloni},
  {Mingo}, {Miskolczi}, {Mooney}, {Nikiel-Wroczy{\'n}ski}, {O'Sullivan},
  {Quinn}, {Reich}, {Roskowi{\'n}ski}, {Rowlinson}, {Savini}, {Saxena},
  {Schwarz}, {Shulevski}, {Sridhar}, {Stacey}, {Urquhart}, {van der Wiel},
  {Varenius}, {Webster}, \& {Wilber}}]{2019A&A...622A...1S}
{Shimwell}, T.~W., {Tasse}, C., {Hardcastle}, M.~J., {et~al.} 2019, \aap, 622,
  A1

\bibitem[{{Smirnov} \& {Tasse}(2015)}]{2015MNRAS.449.2668S}
{Smirnov}, O.~M. \& {Tasse}, C. 2015, \mnras, 449, 2668

\bibitem[{{Struble} \& {Rood}(1991)}]{1991ApJS...77..363S}
{Struble}, M.~F. \& {Rood}, H.~J. 1991, \apjs, 77, 363

\bibitem[{{Szabo} {et~al.}(2011){Szabo}, {Pierpaoli}, {Dong}, {Pipino}, \&
  {Gunn}}]{Szabo_2011}
{Szabo}, T., {Pierpaoli}, E., {Dong}, F., {Pipino}, A., \& {Gunn}, J. 2011,
  \apj, 736, 21

\bibitem[{{Tasse}(2014{\natexlab{a}})}]{2014arXiv1410.8706T}
{Tasse}, C. 2014{\natexlab{a}}, arXiv e-prints, arXiv:1410.8706

\bibitem[{{Tasse}(2014{\natexlab{b}})}]{2014A&A...566A.127T}
{Tasse}, C. 2014{\natexlab{b}}, \aap, 566, A127

\bibitem[{{Tasse} {et~al.}(2018){Tasse}, {Hugo}, {Mirmont}, {Smirnov},
  {Atemkeng}, {Bester}, {Hardcastle}, {Lakhoo}, {Perkins}, \&
  {Shimwell}}]{2018A&A...611A..87T}
{Tasse}, C., {Hugo}, B., {Mirmont}, M., {et~al.} 2018, \aap, 611, A87

\bibitem[{{Tasse} {et~al.}(2021){Tasse}, {Shimwell}, {Hardcastle},
  {O'Sullivan}, {van Weeren}, {Best}, {Bester}, {Hugo}, {Smirnov}, {Sabater},
  {Calistro-Rivera}, {de Gasperin}, {Morabito}, {R{\"o}ttgering}, {Williams},
  {Bonato}, {Bondi}, {Botteon}, {Br{\"u}ggen}, {Brunetti}, {Chy{\.z}y},
  {Garrett}, {G{\"u}rkan}, {Jarvis}, {Kondapally}, {Mandal}, {Prandoni},
  {Repetti}, {Retana-Montenegro}, {Schwarz}, {Shulevski}, \&
  {Wiaux}}]{2021A&A...648A...1T}
{Tasse}, C., {Shimwell}, T., {Hardcastle}, M.~J., {et~al.} 2021, \aap, 648, A1

\bibitem[{{van Diepen} {et~al.}(2018){van Diepen}, {Dijkema}, \&
  {Offringa}}]{2018ascl.soft04003V}
{van Diepen}, G., {Dijkema}, T.~J., \& {Offringa}, A. 2018, {DPPP: Default
  Pre-Processing Pipeline}

\bibitem[{{van Haarlem} {et~al.}(2013){van Haarlem}, {Wise}, {Gunst}, {Heald},
  {McKean}, {Hessels}, {de Bruyn}, {Nijboer}, {Swinbank}, {Fallows},
  {Brentjens}, {Nelles}, {Beck}, {Falcke}, {Fender}, {H{\"o}randel},
  {Koopmans}, {Mann}, {Miley}, {R{\"o}ttgering}, {Stappers}, {Wijers},
  {Zaroubi}, {van den Akker}, {Alexov}, {Anderson}, {Anderson}, {van Ardenne},
  {Arts}, {Asgekar}, {Avruch}, {Batejat}, {B{\"a}hren}, {Bell}, {Bell}, {van
  Bemmel}, {Bennema}, {Bentum}, {Bernardi}, {Best}, {B{\^i}rzan}, {Bonafede},
  {Boonstra}, {Braun}, {Bregman}, {Breitling}, {van de Brink}, {Broderick},
  {Broekema}, {Brouw}, {Br{\"u}ggen}, {Butcher}, {van Cappellen}, {Ciardi},
  {Coenen}, {Conway}, {Coolen}, {Corstanje}, {Damstra}, {Davies}, {Deller},
  {Dettmar}, {van Diepen}, {Dijkstra}, {Donker}, {Doorduin}, {Dromer}, {Drost},
  {van Duin}, {Eisl{\"o}ffel}, {van Enst}, {Ferrari}, {Frieswijk}, {Gankema},
  {Garrett}, {de Gasperin}, {Gerbers}, {de Geus}, {Grie{\ss}meier}, {Grit},
  {Gruppen}, {Hamaker}, {Hassall}, {Hoeft}, {Holties}, {Horneffer}, {van der
  Horst}, {van Houwelingen}, {Huijgen}, {Iacobelli}, {Intema}, {Jackson},
  {Jelic}, {de Jong}, {Juette}, {Kant}, {Karastergiou}, {Koers}, {Kollen},
  {Kondratiev}, {Kooistra}, {Koopman}, {Koster}, {Kuniyoshi}, {Kramer},
  {Kuper}, {Lambropoulos}, {Law}, {van Leeuwen}, {Lemaitre}, {Loose}, {Maat},
  {Macario}, {Markoff}, {Masters}, {McFadden}, {McKay-Bukowski}, {Meijering},
  {Meulman}, {Mevius}, {Middelberg}, {Millenaar}, {Miller-Jones}, {Mohan},
  {Mol}, {Morawietz}, {Morganti}, {Mulcahy}, {Mulder}, {Munk}, {Nieuwenhuis},
  {van Nieuwpoort}, {Noordam}, {Norden}, {Noutsos}, {Offringa}, {Olofsson},
  {Omar}, {Orr{\'u}}, {Overeem}, {Paas}, {Pandey-Pommier}, {Pandey}, {Pizzo},
  {Polatidis}, {Rafferty}, {Rawlings}, {Reich}, {de Reijer}, {Reitsma},
  {Renting}, {Riemers}, {Rol}, {Romein}, {Roosjen}, {Ruiter}, {Scaife}, {van
  der Schaaf}, {Scheers}, {Schellart}, {Schoenmakers}, {Schoonderbeek},
  {Serylak}, {Shulevski}, {Sluman}, {Smirnov}, {Sobey}, {Spreeuw}, {Steinmetz},
  {Sterks}, {Stiepel}, {Stuurwold}, {Tagger}, {Tang}, {Tasse}, {Thomas},
  {Thoudam}, {Toribio}, {van der Tol}, {Usov}, {van Veelen}, {van der Veen},
  {ter Veen}, {Verbiest}, {Vermeulen}, {Vermaas}, {Vocks}, {Vogt}, {de Vos},
  {van der Wal}, {van Weeren}, {Weggemans}, {Weltevrede}, {White}, {Wijnholds},
  {Wilhelmsson}, {Wucknitz}, {Yatawatta}, {Zarka}, {Zensus}, \& {van
  Zwieten}}]{lofar_2013}
{van Haarlem}, M.~P., {Wise}, M.~W., {Gunst}, A.~W., {et~al.} 2013, \aap, 556,
  A2

\bibitem[{{van Weeren} {et~al.}(2019){van Weeren}, {de Gasperin}, {Akamatsu},
  {Br{\"u}ggen}, {Feretti}, {Kang}, {Stroe}, \& {Zandanel}}]{vanWeeren2019}
{van Weeren}, R.~J., {de Gasperin}, F., {Akamatsu}, H., {et~al.} 2019, \ssr,
  215, 16

\bibitem[{{van Weeren} {et~al.}(2010){van Weeren}, {R{\"o}ttgering},
  {Br{\"u}ggen}, \& {Hoeft}}]{2010Sci...330..347V}
{van Weeren}, R.~J., {R{\"o}ttgering}, H. J.~A., {Br{\"u}ggen}, M., \& {Hoeft},
  M. 2010, Science, 330, 347

\bibitem[{{van Weeren} {et~al.}(2012{\natexlab{a}}){van Weeren},
  {R{\"o}ttgering}, {Intema}, {Rudnick}, {Br{\"u}ggen}, {Hoeft}, \&
  {Oonk}}]{2012A&A...546A.124V}
{van Weeren}, R.~J., {R{\"o}ttgering}, H.~J.~A., {Intema}, H.~T., {et~al.}
  2012{\natexlab{a}}, \aap, 546, A124

\bibitem[{{van Weeren} {et~al.}(2012{\natexlab{b}}){van Weeren},
  {R{\"o}ttgering}, {Rafferty}, {Pizzo}, {Bonafede}, {Br{\"u}ggen}, {Brunetti},
  {Ferrari}, {Orr{\`u}}, {Heald}, {McKean}, {Tasse}, {de Gasperin},
  {B{\^\i}rzan}, {van Zwieten}, {van der Tol}, {Shulevski}, {Jackson},
  {Offringa}, {Conway}, {Intema}, {Clarke}, {van Bemmel}, {Miley}, {White},
  {Hoeft}, {Cassano}, {Macario}, {Morganti}, {Wise}, {Horellou}, {Valentijn},
  {Wucknitz}, {Kuijken}, {En{\ss}lin}, {Anderson}, {Asgekar}, {Avruch}, {Beck},
  {Bell}, {Bell}, {Bentum}, {Bernardi}, {Best}, {Boonstra}, {Brentjens}, {van
  de Brink}, {Broderick}, {Brouw}, {Butcher}, {van Cappellen}, {Ciardi},
  {Eisl{\"o}ffel}, {Falcke}, {Fender}, {Garrett}, {Gerbers}, {Gunst}, {van
  Haarlem}, {Hamaker}, {Hassall}, {Hessels}, {Koopmans}, {Kuper}, {van
  Leeuwen}, {Maat}, {Millenaar}, {Munk}, {Nijboer}, {Noordam}, {Pandey},
  {Pandey-Pommier}, {Polatidis}, {Reich}, {Scaife}, {Schoenmakers}, {Sluman},
  {Stappers}, {Steinmetz}, {Swinbank}, {Tagger}, {Tang}, {Vermeulen}, {de Vos},
  \& {van Haarlem}}]{2012A&A...543A..43V}
{van Weeren}, R.~J., {R{\"o}ttgering}, H.~J.~A., {Rafferty}, D.~A., {et~al.}
  2012{\natexlab{b}}, \aap, 543, A43

\bibitem[{{van Weeren} {et~al.}(2020){van Weeren}, {Shimwell}, {Botteon},
  {Brunetti}, {Br{\"u}ggen}, {Boxelaar}, {Cassano}, {Di Gennaro},
  {Andrade-Santos}, {Bonnassieux}, {Bonafede}, {Cuciti}, {Dallacasa}, {de
  Gasperin}, {Gastaldello}, {Hardcastle}, {Hoeft}, {Kraft}, {Mandal},
  {Rossetti}, {R{\"o}ttgering}, {Tasse}, \& {Wilber}}]{2020arXiv201102387V}
{van Weeren}, R.~J., {Shimwell}, T.~W., {Botteon}, A., {et~al.} 2020, arXiv
  e-prints, arXiv:2011.02387

\bibitem[{{van Weeren} {et~al.}(2016){van Weeren}, {Williams}, {Hardcastle},
  {Shimwell}, {Rafferty}, {Sabater}, {Heald}, {Sridhar}, {Dijkema}, {Brunetti},
  {Br{\"u}ggen}, {Andrade-Santos}, {Ogrean}, {R{\"o}ttgering}, {Dawson},
  {Forman}, {de Gasperin}, {Jones}, {Miley}, {Rudnick}, {Sarazin}, {Bonafede},
  {Best}, {B{\^i}rzan}, {Cassano}, {Chy{\.z}y}, {Croston}, {Ensslin},
  {Ferrari}, {Hoeft}, {Horellou}, {Jarvis}, {Kraft}, {Mevius}, {Intema},
  {Murray}, {Orr{\'u}}, {Pizzo}, {Simionescu}, {Stroe}, {van der Tol}, \&
  {White}}]{reinout_2016}
{van Weeren}, R.~J., {Williams}, W.~L., {Hardcastle}, M.~J., {et~al.} 2016,
  \apjs, 223, 2

\bibitem[{{Vikhlinin} {et~al.}(2005){Vikhlinin}, {Markevitch}, {Murray},
  {Jones}, {Forman}, \& {Van Speybroeck}}]{2005ApJ...628..655V}
{Vikhlinin}, A., {Markevitch}, M., {Murray}, S.~S., {et~al.} 2005, \apj, 628,
  655

\bibitem[{{Williams} {et~al.}(2016){Williams}, {van Weeren}, {R{\"o}ttgering},
  {Best}, {Dijkema}, {de Gasperin}, {Hardcastle}, {Heald}, {Prandoni},
  {Sabater}, {Shimwell}, {Tasse}, {van Bemmel}, {Br{\"u}ggen}, {Brunetti},
  {Conway}, {En{\ss}lin}, {Engels}, {Falcke}, {Ferrari}, {Haverkorn},
  {Jackson}, {Jarvis}, {Kapi{\'n}ska}, {Mahony}, {Miley}, {Morabito},
  {Morganti}, {Orr{\'u}}, {Retana-Montenegro}, {Sridhar}, {Toribio}, {White},
  {Wise}, \& {Zwart}}]{williams_2016}
{Williams}, W.~L., {van Weeren}, R.~J., {R{\"o}ttgering}, H.~J.~A., {et~al.}
  2016, \mnras, 460, 2385

\bibitem[{{Wright}(2006)}]{2006PASP..118.1711W}
{Wright}, E.~L. 2006, \pasp, 118, 1711

\bibitem[{{Zhang} {et~al.}(2011){Zhang}, {B{\"o}hringer}, {Finoguenov},
  {Ikebe}, {Matsushita}, {Schuecker}, {Guzzo}, \&
  {Collins}}]{2011A&A...527C...2Z}
{Zhang}, Y.-Y., {B{\"o}hringer}, H., {Finoguenov}, A., {et~al.} 2011, \aap,
  527, C2

\end{thebibliography}

\end{document}